# McFACTS. IV. Electromagnetic Counterparts to AGN-disk-embedded Binary Black Hole Mergers

Emily McPike[1,2], Rosalba Perna[3], K.E. Saavik Ford[1,2,4,5], Barry McKernan[1,2,4,5], Harrison E. Cook[6], Vera Delfavero[7], Miranda McCarthy[1,2], Kaila Nathaniel[8], Jake Postiglione[1,2], Nicolas Posner[9], Varun Pritmani[2], Shawn Ray[1,2], and Richard O'Shaughnessy[8]

[1] Graduate Center, City University of New York, 365 5th Avenue, New York, NY 10016, USA; mcpike.ej@gmail.com
[2] Department of Astrophysics, American Museum of Natural History, New York, NY 10024, USA
[3] Department of Physics and Astronomy, Stony Brook University, Stony Brook, NY 11794, USA
[4] Department of Science, BMCC, City University of New York, New York, NY 10007, USA
[5] Center for Computational Astrophysics, Flatiron Institute, 162 5th Avenue, New York, NY 10010, USA
[6] New Mexico State University, Department of Astronomy, PO Box 30001 MSC 4500, Las Cruces, NM 88003, USA
[7] Canadian Institute for Theoretical Astrophysics, University of Toronto, 60 St George St, Toronto, ON M5S 3H8, Canada
[8] Center for Computational Relativity and Gravitation, Rochester Institute of Technology, Rochester, NY 14623, USA
[9] Data Science Institute, University of Chicago, 5801 S. Ellis Avenue, Chicago, IL 60637, USA
*Received 2026 February 3; revised 2026 June 18; accepted 2026 June 25; published 2026 August 11*

## Abstract

The accretion disks of active galactic nuclei (AGNs) are promising environments for producing binary black hole (BBH) mergers, which have been detected via gravitational waves (GWs) with the LIGO−Virgo−KAGRA (LVK) GW detector network. BBH mergers embedded in AGN disks are unique among GW formation channels in their generic ability to produce electromagnetic (EM) counterparts, via interactions between the merger remnant and the surrounding disk gas (though these are not always observable). While such mergers represent valuable multimessenger sources, the lack of predictive statistical models in existing literature currently limits our ability to select possible EM counterparts with GW detections in archival data and in real time using time-domain surveys such as the Zwicky Transient Facility or LSST. Here we employ the Monte Carlo For AGN Channel Testing and Simulation code (`McFACTS`, https://www.github.com/mcfacts/mcfacts) to predict the bolometric luminosities of jets and shocks associated with LVK-detectable BBH merger remnants in AGN disks. `McFACTS` predicts the distribution of GW observables for an underlying black hole population and disk model. In this work we present a new capability that simultaneously generates the distribution of bolometric EM luminosities corresponding to these predicted GW detections. Our results are consistent with current observational surveys and indicate that (i) migration in dense, Sirko–Goodman-like disks preferentially produces EM counterparts from high-mass ($\mathcal{M} \gtrsim 40\,M_\odot$), high-spin remnants across multiple merger generations and (ii) lower chirp mass mergers are more likely to contribute observable counterparts and with shorter emission breakout times in less dense, Thompson–Quataert–Murray-like disks.

## 1. Introduction

As the LIGO–Virgo–KAGRA (LVK) detector network (T. Akutsu et al. 2020; F. Acernese et al. 2023; E. Capote et al. 2025; S. Soni et al. 2025; The LIGO Scientific Collaboration et al. 2026a, 2026b) continues to frequently detect binary black hole (BBH) mergers, active galactic nuclei (AGNs) remain important environments to study as an efficient channel for BBH formation and merger. Stellar-mass black holes (BHs) from an active galaxies' nuclear star cluster (NSC) can be orbitally coincident with the AGN accretion disk, or captured by the AGN disk as a result of gas dampening and dynamical cooling (K. E. S. Ford & B. McKernan 2026, for a review). Migration in the disk can promote regions of higher number densities of sBHs, promoting the formation of BBHs within the disk (A. Secunda et al. 2020). The relatively high escape velocity, $v_{\rm esc} = \mathcal{O}(10^4\ {\rm km\ s^{-1}}(a/10^3\,R_g))$ ($R_g = GM_{\rm SMBH}/c$), where $a$ is the semimajor axis, imposed by the central supermassive BH (SMBH) makes exiting the disk post-merger difficult, even with relatively strong gravitational-wave (GW) recoil kick velocities imparted after merger ($v_{\rm kick} \leqslant \mathcal{O}(10^3\ {\rm km\ s^{-1}}(a/10^3\,R_g))$; see S. Ray et al. 2026, in preparation). Therefore, the disk can retain the embedded population. Combined, these effects facilitate efficient hierarchical mergers in the disk that can be detected in GWs (B. McKernan et al. 2014; I. Bartos et al. 2017; N. C. Stone et al. 2017; Y. Yang et al. 2019; J. Samsing et al. 2022; K. E. S. Ford & B. McKernan 2026). Repeated mergers produce a population of high-mass and, notably, high-spin BHs, with spins increasingly aligned with the disk and with each other with increasing generation of merger (B. McKernan & K. E. S. Ford 2024). Mergers that happen within this environment constitute the AGN channel (see B. McKernan et al. 2018; H. Tagawa et al. 2020; K. E. S. Ford & B. McKernan 2022; A. Vajpeyi et al. 2022; C. Rowan et al. 2023; G. Fabj & J. Samsing 2024; M. P. Vaccaro et al. 2024; S. H. W. Leong & J. Calderón Bustillo 2025; Y.-J. Li et al.

2026; L. Xue et al. 2025; I. Bartos & Z. Haiman 2026a; Y.-J. Li et al. 2026, and many others for studies that have constrained properties of the AGN channel).

Recent observing runs have substantially increased the GW census. The most recent LVK catalog update, GWTC-5 (The LIGO Scientific Collaboration et al. 2026b), has broadly revealed three populations: (i) most (∼80%) BBHs around $M_1 \sim 10\ M_\odot$, with mostly very low spin ($\chi \sim 0.01$); (ii) a modest-spin ($\chi \sim 0.1$–$0.2$) BBH population around $M_1 \sim 35\ M_\odot$; and (iii) a high-spin ($\chi \gtrsim 0.7$), high-$M_1$ BBH population. The last population is likely attributable to hierarchical mergers and can therefore be tested against Monte Carlo predictions for AGN-channel merger populations (H. Tagawa et al. 2020; H. E. Cook et al. 2025; V. Delfavero et al. 2025a, 2025b; B. McKernan et al. 2025). The fourth GW transient catalog, The LIGO Scientific Collaboration et al. (2026a), included the most massive merger event yet detected by LVK, GW231123 (The LIGO Scientific Collaboration et al. 2025), and hinted at a subpopulation of aligned-spin mergers (Y.-J. Li et al. 2026). Massive (upper mass gap) BHs, high spins, and a bias toward aligned-spin mergers can be well explained by hierarchical merger scenarios, including the AGN channel (V. Delfavero et al. 2025b). The AGN disk channel produces a subpopulation of events with distinctive properties (asymmetric binaries; high-mass, high-spin-aligned binaries), and this characteristic signature can be used to disentangle the AGN disk contribution in the currently observed population (A. Vajpeyi et al. 2022; L.-G. Zhu & X. Chen 2025); these and similar preliminary analyses provide increasing evidence that this fraction is nonzero. Still, the exact fraction of such events ($f_{\rm BBH,AGN}$) remains an open question. If AGN disks are commonly geometrically thin, dense, and of moderate radial size ($\lesssim$0.1 pc) and lifetime ($\lesssim$Myr), earlier estimates spanned $f_{\rm BBH,AGN} \sim 5\%$–$80\%$ (e.g., K. E. S. Ford & B. McKernan 2022; L.-G. Zhu & X. Chen 2025). However, the three broad populations from GWTC-5 discussed above suggest $f_{\rm BBH,AGN} \sim \mathcal{O}(1\%-10\%)$ (I. Bartos & Z. Haiman 2026b). Nevertheless, if AGN disks are mostly very small radially ($\lesssim$0.01 pc), or very low in density and geometrically thick, $f_{\rm BBH,AGN}$ could be very small (Y. Tomar et al. 2026). Merger components and remnant characteristics have been studied in association with various BBH formation channels (see M. Mapelli 2021, for a review). For the AGN channel, given that the AGN accretion disk is gas-rich, the embedded BHs must interact with the surrounding disk gas. Angular momentum transfer via gas torques from the disk can align and increase the spin of embedded BHs via accretion and mergers (K. E. S. Ford & B. McKernan 2026). Gas transfers angular momentum via accretion onto embedded BHs (I. Bartos & Z. Haiman 2026a). Repeated mergers in AGN disks can make intermediate-mass BHs (IMBHs; $M > 10^2\ M_\odot$) and, if the repeated mergers occur prograde in the same plane, can also boost the merged BH spin to high values ($\chi \sim 0.7$–$0.95$) depending on merger generation (V. Delfavero et al. 2025b; B. McKernan et al. 2025). It is important to acknowledge that other channels of BBH formation can also produce high-mass mergers, e.g., in the isolated binary scenario (K. Belczynski et al. 2010; S. E. de Mink & I. Mandel 2016) or via multiple mergers in stellar clusters (e.g., contributions from NSCs, globular clusters; C. L. Rodriguez et al. 2016; J. Hong et al. 2018; K. Kritos et al. 2023; C. S. Ye & M. Fishbach 2024), although merged BH spins in dynamical, gas-free environments are likely capped at (dimensionless spin parameter) $\chi \sim 0.8$ (K. Kritos et al. 2023).

The AGN channel is the only BBH production pathway where mergers practically always occur within a gas-rich environment and can thus easily produce electromagnetic (EM) counterparts. Post-merger, a remnant BH interacts with surrounding AGN disk gas. Rapid accretion onto embedded remnants can produce potentially luminous EM counterparts to GWs (M. J. Graham et al. 2020; H. Tagawa et al. 2023; K. Chen & Z.-G. Dai 2024; J. C. Rodríguez-Ramírez et al. 2025). Two general luminosity-generating mechanisms have been proposed: (i) the change in orbit from the GW recoil kick imparted to the remnant at merger will dissipate energy into the disk over time and potentially generate luminous shocks (B. McKernan et al. 2019), and (ii) the kicked remnant rapidly accretes gas from the disk, which, when combined with a high BH spin ($a \sim 0.5$–$0.9$), could trigger the remnant BH to drive a jet (e.g., M. J. Graham et al. 2020; Y. Kim & E. R. Most 2025), alongside more specific outflow models suggested by J. C. Rodríguez-Ramírez et al. (2024). Thermal and nonthermal emission from jets produced by BH remnants has been proposed to be observable in UV/optical and X-ray/gamma-ray radiation (H. Tagawa et al. 2023; K. Chen & Z.-G. Dai 2024; J. C. Rodríguez-Ramírez et al. 2024; H.-H. Zhang et al. 2024; J. C. Rodríguez-Ramírez et al. 2025). Semianalytic models have been constructed to study both the thermal and nonthermal emission for conditions expected for embedded BH remnants (K. Chen & Z.-G. Dai 2025). Such models reproduce observations of candidate GW counterparts by assuming that remnants can attain various accretion rates up to 15 times the Bondi–Hoyle–Lyttleton (BHL) mass accretion rate ($\dot{M}_{\rm Bondi}$) (e.g., H. Tagawa et al. 2023), although recent general relativistic magnetohydrodynamic (GRMHD) simulations of embedded BHs in AGN disks suggest a limiting accretion rate of $\sim 0.1\dot{M}_{\rm Bondi}$ (A. J. Dittmann et al. 2024).

However, there are challenges to effectively observing potential EM counterparts to BBH mergers. AGNs (here spanning Seyfert AGNs and quasars) are bright ($L_{\rm AGN} \sim 10^{42}-10^{47}$ erg s$^{-1}$) and variable (M.-H. Ulrich et al. 1997). The driving mechanisms behind AGN variability (e.g., turbulence, magnetorotational instability, coronal heating, variable accretion rates; Y.-F. Jiang et al. 2019; S. Hagen et al. 2024; O. Blaes et al. 2025) remain poorly constrained. Thus, the false positives in AGN light curves for EM counterparts to BBH mergers are poorly understood. Even if a sufficiently luminous EM counterpart is produced, an optically thick disk could prevent counterpart emission from escaping the disk rapidly, making it effectively undetectable by the time it reaches the disk photosphere. Even in the subset of cases where counterpart emission is able to escape the disk promptly (such as in the case of jet breakout), the flare might be dim and indistinguishable from a bright, variable AGN disk.

Additionally, geometric constraints limit detectability. Assuming roughly equal numbers of Type I (face-on) and Type II (edge-on) AGNs, with roughly equal numbers for each, a reasonable rule of thumb is that the fraction of all signatures originating in Type II AGNs are not detectable. Up to half of those in Type I AGNs may also be missed if breakout occurs on the other side of the disk. Thus, geometrically only $\mathcal{O}(1/4)$ of all breakout EM signatures are *in principle* detectable.

Actually following up GW detections with EM observatories poses an additional problem. Poor localization of the GW signal on the sky and in redshift ($z$) by LVK poses yet another obstacle to observing EM counterparts. Large LVK error volumes can prevent effective EM follow-up to GW events given the large number of AGNs on such extended regions of the sky. We hope that the very recent release of chirp mass ($\mathcal{M}_{\rm chirp}$) bins by LVK will help with EM follow-up in the proposed observing run later in 2026.

Despite observational challenges, efforts are being made, both archivally (e.g., V. Connaughton et al. 2016; M. J. Graham et al. 2020, 2023) and in real time (e.g., T. Cabrera et al. 2024; P. Darc et al. 2025), to search for EM counterparts to LVK-detected BBH mergers. Campaigns span a broad range of wavelengths. In the optical band, wide-field surveys with the Zwicky Transient Facility (ZTF) have been the most systematic, targeting AGN flares coincident with O3 BBH events (M. J. Graham et al. 2023) and yielding up to seven ZTF AGN flares that matched nine GW events across the run, though reanalysis with extended baselines reduced confident candidate AGN–GW pairs to two (L. He et al. 2025). While ZTF follow-up observations during O4a and O4b identified four flare candidates (H. Bommireddy et al. 2026), none exhibited the asymmetric broad-line profile expected from an off-center explosive event, which is considered a "smoking-gun" signature of a genuine counterpart[10] (M. J. Graham et al. 2023; T. Cabrera et al. 2024). Targeted real-time campaigns with DECam (T. Cabrera et al. 2024) and STEP (A. Santos et al. 2024; P. Darc et al. 2025) have extended coverage to the southern sky and provide spectroscopic follow-up, with T. Cabrera et al. (2024) finding one candidate counterpart to S230922g with hints of asymmetric broad-line responses. Broadly, optical campaigns struggle to identify unambiguous counterparts amid AGN variability and large GW localization volumes. At high energies, two GRB candidates have been proposed, first in O1 (B. P. Abbott et al. 2016; V. Connaughton et al. 2016, 2018) and most recently in O4a (J. DeLaunay et al. 2024; S.-R. Zhang et al. 2026), the latter detected with an X-ray afterglow. Other coordinated searches with Fermi/Gamma-ray Burst Monitor, Swift, and other instruments have not revealed BBH EM counterpart candidates outside of the proposed counterpart to S241125n.[11] A relatively limited, putative sample of $\gamma$-ray and X-ray candidate counterparts could place meaningful upper limits on prompt high-energy emission (R. A. J. Eyles-Ferris et al. 2025; H. Bommireddy et al. 2026).

However, uncertainty in the physical mechanisms driving AGN variability (e.g., O. Blaes et al. 2025), as well as in the nature of potential EM counterparts (e.g., H. Tagawa et al. 2024), has made robust GW–EM associations difficult to establish (e.g., A. Palmese et al. 2021; N. Veronesi et al. 2025). Nevertheless, considerable effort has been devoted to identifying plausible counterparts and assessing associations statistically. For example, T. Cabrera et al. (2026) analyzed AGN-channel candidate mergers in the LVK catalog and inferred that only $\sim$3% of GW events within the ZTF footprint have genuine observed EM counterparts at the 90% confidence level. This low fraction likely reflects both the intrinsic rarity of detectable EM emission and the limited sky coverage of ZTF.

With a limited candidate sample size and uncertainties as outlined above, it is clear that detailed, physically motivated emission models for EM counterparts with anticipated timescales of emission are highly desirable and will provide important guidance for current and future follow-up campaigns.

The payoff for successfully observing EM counterparts to BBH mergers in AGN is significant. Multimessenger counterparts are crucial to strongly and independently constraining $H_0$. With a confident EM counterpart to a BBH merger, a redshift measurement from the flare combined with the luminosity distance from the GW can provide a direct constraint on $H_0$ (see, e.g., S. Nissanke et al. 2013; A. Palmese & S. Mastrogiovanni 2026 for a review). Efforts have been made both to bolster the statistical connection between proposed flares and to constrain $H_0$ with current proposed catalogs of putative EM counterparts (T. Cabrera et al. 2026). These studies have reported $H_0$ to an accuracy of $\sim$7% (A. Palmese & S. Mastrogiovanni 2026) and are consistent with measurements of $H_0$ from A. G. Riess et al. (2019) and Planck Collaboration et al. (2020). In the upcoming decade, we expect to see a large increase in the statistical BBH merger sample. Therefore, it is essential to model EM counterparts to BBH mergers to better our EM follow-up methods in anticipation of the future wealth of data.

There has yet to be a fine-tunable population synthesis model for these LVK AGN-channel EM counterpart events available for testing emission models. Here we present the first population synthesis study of EM counterparts to AGN-channel BBHs. More specifically, this work introduces new capabilities in `McFACTS` (H. E. Cook et al. 2025; V. Delfavero et al. 2025a; B. McKernan et al. 2025) to self-consistently generate the distributions of the bolometric jet and shock luminosities as EM counterparts to LVK-detectable GW events produced by BBH mergers in the AGN channel. These results also provide constraints on the merger properties and characteristic timescales expected for BBHs in AGN environments.

## 2. `McFACTS v.0.4.0`

`McFACTS` is a 1D, open-source rapid population synthesis code for dynamic compact binary formation channels in AGN disks (H. E. Cook et al. 2025; V. Delfavero et al. 2025a; B. McKernan et al. 2025, hereafter Papers I, II, and III, respectively). Here we outline the released `v.0.4.1` of the `McFACTS` code and highlight important updates from the previously published version `v.0.3.0`. [12]

First, the code is now significantly faster, due to improvements in code efficiency. `McFACTS v.0.4.1` is much faster than `v.0.3.0`, with a grid of 100 AGN galaxy iterations running in 10–20 s on an Apple M3 (Model Identifier: Mac15,12). This test was carried out with default settings (see Table 1), for 70 time steps of duration $10^4$ yr. This is nearly an order-of-magnitude improvement in the computation time over `v.0.3.0`.

[10] ZTF coverage is limited to the northern hemisphere, leaving a substantial fraction of potential counterparts unobservable.

[11] D. Lazzati et al. (2023) proposed GRB 191019A to be produced by a binary neutron star merger within an AGN disk; however, since the distance of this GRB placed it outside of the LIGO horizon, this event was not simultaneously detected via GWs.

[12] The reader can find full documentation at https://github.com/mcfacts.

**Table 1**
Fiducial Model Parameters for the SMBH, AGN Disk, and NSC in paper_4/pem_sg_default.ini

| Parameter | Default Value | Units |
|---|---|---|
| smbh_mass | $10^8$ | $M_\odot$ |
| disk_model_name | sirko_goodman | |
| flag_use_pagn | 1 | |
| flag_use_surrogate | −1 | |
| flag_use_spin_check | 0 | |
| flag_phenom_turb | 0 | |
| phenom_turb_centroid | 0 | |
| phenom_turb_std_dev | 1.0 | |
| disk_radius_trap | 700 | $r_{g,\ \mathrm{SMBH}}$ |
| disk_radius_outer | $5 \times 10^4$ | $r_{g,\ \mathrm{SMBH}}$ |
| disk_radius_max_pc | 0 | |
| disk_alpha_viscosity | 0.01 | |
| inner_disk_outer_radius | 50 | $r_{g,\ \mathrm{SMBH}}$ |
| disk_inner_stable_circ_orb | 6 | $r_{g,\ \mathrm{SMBH}}$ |
| disk_aspect_ratio_avg | 0.03 | |
| nsc_radius_outer | 5 | pc |
| nsc_radius_crit | 0.25 | pc |
| nsc_mass | $3 \times 10^7$ | $M_\odot$ |
| nsc_ratio_bh_num_star_num | $10^{-3}$ | |
| nsc_ratio_bh_mass_star_mass | 10 | |
| nsc_density_index_inner | 1.75 | |
| nsc_density_index_outer | 2.5 | |
| nsc_spheroid_normalization | 1 | |
| nsc_imf_bh_mode | 10 | $M_\odot$ |
| nsc_imf_bh_max | 40 | $M_\odot$ |
| nsc_imf_powerlaw_index | 2 | |
| nsc_imf_bh_method | default | |
| nsc_bh_spin_dist_mu | 0 | |
| nsc_bh_spin_dist_sigma | 0.1 | |
| nsc_imf_powerlaw_index | 2 | |
| disk_bh_torque_condition | 0.1 | |
| disk_bh_eddington_ratio | 1.0 | |
| disk_bh_orb_ecc_max_init | 0.9 | |
| disk_radius_capture_outer | $10^3$ | $r_{g,\ \mathrm{SMBH}}$ |
| disk_bh_pro_orb_ecc_crit | 0.01 | |
| mass_pile_up | 35 | $M_\odot$ |
| timestep_duration_yr | $10^4$ | yr |
| timestep_num | 70 | |
| capture_time_yr | $10^5$ | yr |
| galaxy_num | 100 | |
| delta_energy_strong_mu | 0.1 | |
| delta_energy_strong_sigma | 0.02 | |
| fraction_bin_retro | 0.0 | |
| flag_thermal_feedback | 1 | |
| flag_orb_ecc_damping | 1 | |
| flag_dynamic_enc | 1 | |
| flag_torque_prescription | jimenez_masset | |
| harden_energy_delta_mu | 0.9 | |
| harden_energy_delta_sigma | 0.025 | |

**Note.** Star parameters are not included since stars are turned off by default in v.0.4.0.

As in Papers I–III, by default we assume that each galaxy has an SMBH of mass $M_{\mathrm{SMBH}} = 10^8\ M_\odot$ accreting at 10% of the Eddington mass accretion rate, thus yielding a bolometric AGN luminosity of $L_{\mathrm{AGN}} \sim 10^{45}\ \mathrm{erg\ s^{-1}}$. We assume by default that the active episode lasts 0.7 Myr.

We compute the radial structure of two disk models by making use of the pAGN package (D. Gangardt et al. 2024): Sirko–Goodman (SG; our fiducial disk model choice; E. Sirko & J. Goodman 2003) and Thompson–Quataert–Murray (TQM; T. A. Thompson et al. 2005). For each model, we adopt the fiducial values described in D. Gangardt et al. (2024). By default, we restrict our analysis to the radial interval $[6,\ 5 \times 10^4]\ R_g$, where $R_g = GM_{\mathrm{SMBH}}/c^2$ is the gravitational radius. The inner disk boundary is set by the innermost stable circular orbit, but McFACTS categorizes any BH within $50R_g$ as an inner disk object, tracked separately to make predictions for extreme mass ratio inspirals (EMRIs; K. E. S. Ford et al. 2026, in preparation). The outer-disk boundary is set both to isolate the region in which migration will contribute to hierarchical mergers and to avoid the far outer-disk regime where both disk models become uncertain owing to their respective treatments of the outer boundary condition.

The choice of outer radius is particularly consequential for the TQM disk model. In TQM, the disk is regulated by star formation feedback that maintains marginal gravitational stability in the outer disk, causing the global disk structure to depend sensitively on the adopted outer boundary conditions and inflow rate. As a result, changes to the outer radius can modify the properties of the disk over a broad range of radii. By contrast, the SG disk model is constructed primarily from local viscous and opacity prescriptions, making its inner structure comparatively less sensitive to the precise choice of outer boundary. Our adopted outer radius therefore enables a more controlled comparison between the two disk models. Following previous studies (J. M. Bellovary et al. 2016), we assume that these disks have migration traps at $R_{\mathrm{mig,\ SG}} = 700R_g$ and $R_{\mathrm{mig,\ TQM}} = 500R_g$, respectively.

Most default model assumptions from Paper I still apply (see that paper for detailed discussion of model choices). We have improved overall code performance compared to v.0.3.0, including a bug fix to our initial eccentricity prescription, which circularized orbits more slowly than it should have. As a result, we have changed default values of disk_bh_orb_ecc_max_init from 0.3 to 0.9 and timestep_num from 50 to 70, respectively (see Table 1), allowing for a more dynamically excited initial BH population on average over a slightly longer default disk lifetime (0.7 Myr). Also new in v.0.4.0, we now evolve the spin vectors for the primary and secondary BHs in the time prior to merger using the precession module (D. Gerosa & M. Kesden 2016), so the treatment of the final BBH merger stage (including kicks) is more accurate (see also S. Ray et al. 2026, in preparation). We have also added a phenomenological turbulence model to migration, which is off in default (flag_phenom_turb=0), but which can be turned on by users (see Section 3.5 below).

## 3. Methods

In this section we outline the mechanisms for generating radiation from interactions between a BBH remnant and the AGN disk and introduce the models we use to compute the luminosities from these interactions, as well as the parameters we vary in the McFACTS code to probe the luminosity distribution of potential EM counterparts generated by the code.

### 3.1. Shock Luminosity

A GW kick perturbs the binary's original orbit (Y. Zlochower et al. 2011), resulting in the remnant's final velocity being a combination of its Keplerian velocity and the kick velocity. As the remnant moves along its new orbit within the disk, gas previously within its Hill sphere spreads into a bow-shaped structure surrounding it. This original Hill sphere gas collides with a comparable mass of accretion disk gas, and we parameterize the bolometric shock luminosity $L_{\rm shock} = E_{\rm shock}/t_{\rm shock}$ following B. McKernan et al. (2019), where

$$t_{\rm shock} = \mathcal{O}(6\ {\rm months}) \times \left(\frac{R_{\rm H}}{3R_g}\right)\left(\frac{v_{\rm kick}}{100\ {\rm km\ s^{-1}}}\right)^{-1} \quad (1)$$

is the duration of the shock, with $v_{\rm kick}$ the GW kick velocity imparted to the post-merger remnant. The energy dissipated ($E_{\rm shock}$) is parameterized as

$$E_{\rm shock} = 10^{47} {\rm erg}\left(\frac{M_{\rm H}}{1\ M_\odot}\right)\left(\frac{v_{\rm kick}}{100\ {\rm km\ s^{-1}}}\right)^2, \quad (2)$$

with $M_{\rm H}$ the mass of gas inside the Hill sphere of radius $R_{\rm H}$, so

$$L_{\rm shock} \sim 10^{40}\ {\rm erg\ s^{-1}}\left(\frac{M_{\rm H}}{1\ M_\odot}\right)\left(\frac{v_{\rm kick}}{10^2\ {\rm km\ s^{-1}}}\right)^3\left(\frac{R_{\rm H}}{3R_g}\right)^{-1}, \quad (3)$$

where $M_{\rm H} = \rho V_{\rm H}$ is the mass of gas contained within the volume ($V_{\rm H}$) of the remnant's Hill sphere with local average gas density $\rho$, $v_{\rm kick}$ is the kick velocity of the merged BH, and $R_{\rm H} = r_{\rm BH}(M_{\rm BH}/M_{\rm SMBH})^{1/3}$ is the Hill sphere radius of the kicked BH of mass $M_{\rm BH}$ at AGN disk radius $r_{\rm BH}$. For each merger remnant in `McFACTS`, we estimate $L_{\rm shock}$ according to Equation (3).

### 3.2. Jet Luminosity

Jets can be launched after BBH mergers if the circumbinary environment supplies sufficient gas for rapid post-merger accretion, enabling the remnant BH to enter a magnetically arrested disk (MAD) state (A. Tchekhovskoy et al. 2011). In such settings, gas inflow from the circumbinary disk or from fallback material that survives the coalescence can feed the remnant at near- or super-Eddington rates, promoting the accumulation and amplification of magnetic flux. The post-merger BH can then compress and advect this flux to build the large-scale poloidal magnetic field required for a Blandford–Znajek (BZ) outflow (R. D. Blandford & R. L. Znajek 1977), provided that its spin is high enough to supply the necessary rotational energy. The transient accretion flow that forms immediately after the merger naturally evolves toward an MAD configuration under these conditions, allowing a relativistic jet to be launched.

Jet power ($L_{\rm jet}$) depends on the accumulation of magnetic flux onto the BH and on the efficiency of energy extraction via the BZ mechanism, which we parameterize as

$$L_{\rm jet} = \eta_{\rm jet}\,\dot{M}c^2\,, \quad (4)$$

where the efficiency parameter $\eta_{\rm jet}$ depends on the BH spin as $\propto a^2$ (with small corrections for high values of $a$; A. Tchekhovskoy et al. 2011) and on the strength with which magnetic flux is transported to the accreting BH via turbulent accretion. GRMHD simulations by A. Tchekhovskoy et al. (2011) have shown $\eta_{\rm jet} > 1$, implying that spinning BHs yield $L_{\rm jet}$ with energies greater than the rest-mass energy supplied by accretion, since the jet taps the BH rotational energy (up to $\sim 0.4 M_{\rm BH}c^2$ for maximal prograde spin).

Following the merger, the remnant BH interacts dynamically and (magneto)hydrodynamically with surrounding disk gas. For a BH of mass $M_{\rm BH}$ moving at relative velocity $v_{\rm rel}$ through gas of density $\rho$ and sound speed $c_s$, the BHL accretion rate is

$$\dot{M}_{\rm BHL} \simeq \frac{4\pi G^2 M_{\rm BH}^2 \rho}{(v_{\rm rel}^2 + c_s^2)^{3/2}}, \quad (5)$$

where $v_{\rm rel} \sim v_{\rm kick}$ and $\dot{M}_{\rm BHL}$ can greatly exceed the Eddington limit ($\dot{M}_{\rm Edd}$) in large parts of our model disk.

While $\dot{M}_{\rm BHL}$ offers a useful order-of-magnitude estimate for accretion onto the remnant BH, numerous studies demonstrate that significant deviations can arise owing to various physical effects. C.-L. Jiao et al. (2025) argue that differential rotation in an AGN disk implies substantial angular momentum in the surrounding disk gas, necessarily driving accretion via a disklike flow rather than a Bondi-type flow with $\dot{M} \ll \dot{M}_{\rm BHL}$. The accumulation of magnetic flux on horizon scales can also reduce $\dot{M}$ (H. Cho et al. 2023). Conversely, shocks due to large $v_{\rm rel}$ can enhance $\dot{M} \sim 10\dot{M}_{\rm BHL}$ (H. Tagawa et al. 2023).

Recent GRMHD studies of jet launching during BHL accretion find $\dot{M} = f_{\rm acc}\dot{M}_{\rm BHL}$ with $f_{\rm acc} \approx 0.05$, with high jet efficiency $\eta_{\rm jet} \approx 2$–3, for BH spins of $a = 0.9$ (N. Kaaz et al. 2023; Y. Kim & E. R. Most 2025). We follow these numerical results by parameterizing $L_{\rm jet}$ as

$$L_{\rm jet} \sim 0.1\left(\frac{f_{\rm acc}\,\eta_{\rm jet}}{0.1}\right)\left(\frac{a}{0.9}\right)^2 \dot{M}_{\rm BHL}\,c^2 \quad (6)$$

and calculating this quantity for all BBH mergers in `McFACTS`.[13] Equation (6) represents an empirical relation trying to capture the dependence of the jet luminosity both on the accreted mass and on the BH spin: it should be considered only as a guide, potentially susceptible to rescaling. Whether a jet can be launched in the first place depends on a combination of BH spin, magnetic flux, and accretion power. Both numerical simulations and observational evidence point to the ubiquitous occurrence of jets from accreting BHs when the accretion rate approaches or exceeds the Eddington rate (e.g., A. J. Goodwin & A. Mummery 2026). These conditions are practically always met at the high densities where the bulk of the mergers occur in our AGN disk simulations. Jets in N. Kaaz et al. (2023) and Y. Kim & E. R. Most (2025) are sporadic, with an active phase lasting approximately $\sim 24{,}000 R_g/c \sim 10^7\ {\rm s}\ (M_{\rm SMBH}/10^8\ M_\odot)$, which we use as our fiducial $T_{\rm jet}$ here, a choice that is well supported by similar simulations (A. Lalakos et al. 2025) and consistent with current flare candidates (H. Bommireddy et al. 2026).

[13] We remark that this enhanced accretion is only assumed to last for the jet lifetime $T_{\rm jet}$. After the engine turns off, the remnant is assumed to simply accrete at the Eddington rate. We assume that the mass accreted during the hyperaccreting phase with an active jet is negligible ($\dot{M}_{\rm BHL}\,T_{\rm jet} \ll M_{\rm BH}$).

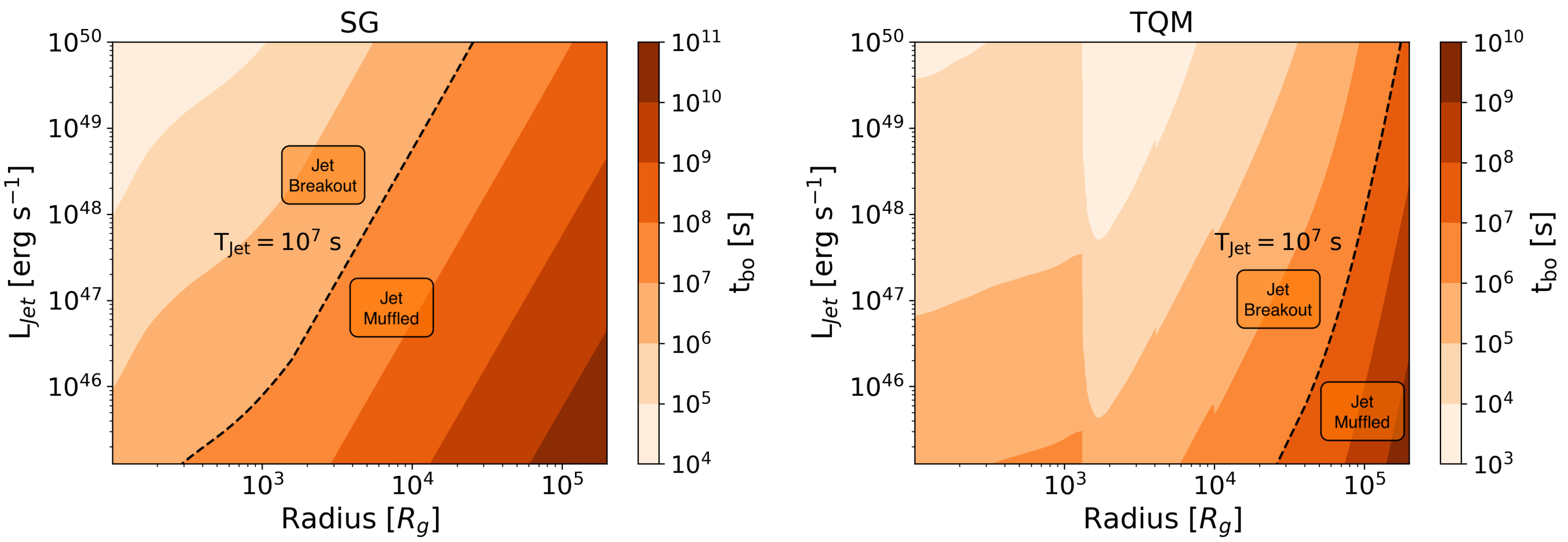

**Figure 1.** Jet breakout time for representative luminosities vs. remnant location. Regions of the disk where, for a given luminosity, a jet is able to break out of the disk before the engine turns off ($t_{\rm bo} < T_{\rm jet}$). Color shades indicate different values of $t_{\rm bo}$. Our fiducial jet lifetime $T_{\rm jet} = 10^7$ s is marked as a dashed black line. Any combination of $L_{\rm jet}$ and radius lying to the left of the dotted line will break out; those to the right are muffled, escaping the disk on the photon diffusion timescale. Left: SG disk model. Right: TQM disk model.

### *3.2.1. Observability of the Jet Emission*

Once a jet is launched by merging BHs within an AGN disk, the detectability of its radiation remains challenging owing to the extreme densities of the surrounding medium—even under the most favorable conditions of a face-on viewing angle. Fundamentally, a jet's observability is dependent on whether it can successfully break out of the disk while it is still active and radiating, thereby allowing its emission to freely escape toward the observer. Here we derive a simple criterion to assess whether the jet is capable of emerging from the disk.

Given a jet of luminosity $L_{\rm jet}$, let $\theta_{\rm jet}$ be its half-opening angle and $T_{\rm jet}$ its duration with. Let $\rho$ be the disk density and $H$ be the disk scale height at a given binary's location. The jet can successfully emerge from the disk ("breakout") if the jet head reaches the disk surface within the engine lifetime ($\mathcal{O}(10^7{\rm s})$).

The jet-head cross section is approximately

$$\sigma \approx \pi(\theta_{\rm jet} H)^2. \tag{7}$$

A convenient dimensionless measure of jet strength is then provided by the dimensionless parameter (C. D. Matzner 2003; O. Bromberg et al. 2011)

$$\tilde{L} \equiv \frac{L_{\rm jet}}{\sigma\ \rho\ c^3} = \frac{L_{\rm jet}}{\pi(\theta_{\rm jet} H)^2\ \rho\ c^3}. \tag{8}$$

This parameter measures the ratio between the energy density of the jet and the rest-mass energy density of the surrounding medium at the jet-head location.

For a relativistic jet expanding in a nonrelativistic medium with a strong reverse shock, the velocity of the jet head can then be approximated via the expression (C. D. Matzner 2003)

$$\beta_h \equiv \frac{v_h}{c} \simeq \frac{\tilde{L}^{1/2}}{1 + \tilde{L}^{1/2}}. \tag{9}$$

The breakout time $t_{\rm bo}$, i.e., the time required for the jet head to traverse the vertical extent $H$ of the disk, is then given by

$$t_{\rm bo} \simeq \frac{H}{v_h} = \frac{H}{\beta_h c}. \tag{10}$$

If the jet head reaches the surface of the disk before the power ceases, the jet can successfully emerge from the disk. The condition for successful breakout is therefore simply

$$T_{\rm jet} > t_{\rm bo} \quad \Rightarrow \quad \text{Jet breakout}. \tag{11}$$

If $T_{\rm jet} < t_{\rm bo}$, the jet stalls and becomes choked, depositing its energy into a hot cocoon rather than producing a directly observable relativistic outflow.

Figure 1 displays the regions in both disk models from which breakout can occur, as a function of the active-phase duration $T_{\rm jet}$, for several representative jet luminosities ($L_{\rm jet}$) and a reference jet opening angle of $\theta_{\rm jet} = 10°$. The dashed line indicates our fiducial jet duration $T_{\rm jet} = 10^7$ s; below this line, the jet is choked before it can emerge from the disk. Since $t_{\rm bo}$ depends on both the density $\rho$ and the scale height $H$, Figure 1 demonstrates that breakout is highly sensitive to both the disk model and the jet lifetime. Note that this simple analysis assumes that the jet travels along the direction perpendicular to the disk plane. If it is instead oriented at some angle $\theta_{\rm obs}$, the jet will have to travel a longer distance of $H/\cos\theta_{\rm obs}$. This will result in slight variations for the jet breakout with respect to what is shown in Figure 1.

If the jet fails to break out of the disk before the engine shuts off, the likelihood of detecting any emission from it before it stalls depends critically on the optical depth between the jet head and the disk photosphere. This is an inherently time-dependent problem, requiring the evolution of the jet within the surrounding medium to be tracked and depending sensitively on the jet's physical properties (see, e.g., R. Perna et al. 2021; J.-P. Zhu et al. 2021; Y.-H. Wang et al. 2022; H. D. Kang et al. 2025 for the specific case of GRB jets), as well as the presence of outflow cavities before the merger (S. S. Kimura et al. 2021). Given that the characteristics of jets launched by post-merger BHs remain largely unconstrained, we adopt a conservative approach and estimate the maximum possible distance to the photosphere from the disk midplane. If the optical depth to Thomson scattering satisfies $\tau \gtrsim 1$, photons escape on the corresponding diffusive timescale,

$$t_{\rm diff} \approx \frac{H\tau}{c}\,, \qquad \text{where} \qquad \tau \approx n\sigma_{\rm T} H\,, \tag{12}$$

where $n = \rho/m_p$, with $m_p$ the proton mass, and $\sigma_T$ is the Thomson cross section. The emerging luminosity can then be estimated as

$$L_{\rm diff} \sim L_{\rm jet} \frac{T_{\rm jet}}{t_{\rm diff}} \frac{\Omega_{\rm jet}}{4\pi}, \tag{13}$$

where $\Omega_{\rm jet}$ represents the solid angle within which the intrinsic emission is beamed; for small angles, $\Omega_{\rm jet} \simeq \pi\theta_{\rm jet}^2$. While we use the above equations in our Monte Carlo simulations, we note that adopting the full scale height $H$ in Equation (12) should be regarded as an upper limit. This choice implicitly assumes that the jet is launched from the midplane and that its radiation always emerges from that same location. In practice, as the jet propagates through the disk, the optical depth between the jet head and the disk photosphere decreases with time (see, e.g., R. Perna et al. 2021; M. Ray et al. 2023, for quantitative analyses).

### 3.3. Disk Lifetime

The lifetime of AGN disks ($\tau_{\rm AGN}$) is widely debated. $\tau_{\rm AGN}$ is thought to lie within the range of ~0.1 Myr to $\mathcal{O}$(10 Myr) (e.g., P. Martini 2004; A. King & C. Nixon 2015; K. Schawinski et al. 2015). The lifetime of the disk impacts how long embedded BHs can continue the chain of hierarchical mergers. Shorter-lived disks may not allow for efficient hierarchical mergers. Longer-lived disks ($\geqslant$2 Myr) could become inefficient as well, due to consumption of the initial population of BHs (Paper III). Considering these dependencies, here we adopt a default value $\tau_{\rm AGN} = 0.7$ Myr but also explore the effect of both shorter ($\tau_{\rm AGN} = 0.25$ Myr) and longer ($\tau_{\rm AGN} = 1$ Myr) disk lifetimes.

### 3.4. Disk Size

The mechanism by which the outer disk is supported is largely unknown, though promising suggestions are radiation pressure by accreting BHs (S. Gilbaum & N. C. Stone 2022; M. Epstein-Martin et al. 2025) and magnetic pressure (P. F. Hopkins et al. 2024; H. J. Gerling-Dunsmore et al. 2026). It also remains unclear whether the accretion disk has a physical connection to the dusty, obscuring torus. Thus, the outer radius of the disk ($R_{\rm outer}$) remains uncertain. The disk size ($\propto R_{\rm outer}^2$) and scale height ($H$) directly determine the number of orbits of compact objects and stars intersecting and captured by the AGN disk from the interacting NSC population. As $R_{\rm outer}$ increases, the number of captured objects rises. However, as NSCs typically have higher densities toward small radii, if $R_{\rm outer}$ grows too much, the average mutual separation between captured BHs can increase, lowering the rate of encounters that form and merge BBHs in AGNs.

TQM disk model physics at small radii is highly sensitive to differing choices of $R_{\rm outer}$. Thus, we exclude this disk model from the runs in which we vary $R_{\rm outer}$ in order to preserve physical consistency.

### 3.5. Phenomenological Disk Turbulence Prescription

By default McFACTS assumes that gas torques yield type I migration of circularized ($e \leqslant e_{\rm crit} = 0.01$) embedded BHs (see Paper I). To test the effect of disk turbulence on migration, we recast the migration distance over a time step as $D(1 + \delta)$, where $D$ is the default Type I migration distance calculated by McFACTS and $\delta$ is an additional stochastic distance component drawn from a Gaussian distribution. The mean and variance of the Gaussian distribution (see Table 1) are user defined, allowing the magnitude and directionality of turbulence-induced fluctuations to be tested. Effectively, increasing the magnitude of $\delta$ introduces a random-walk component to the standard mass-weighted Type I migration, thereby adding an increasing random component to the BBH pairings that result from migration.

### 3.6. BH Initial Mass Function Models

We test several BH initial mass function (IMF) models. By default we draw BHs from a Pareto distribution function ($P(M) \propto M^{-\gamma}$) spanning [10, 40] $M_\odot$, with a Gaussian bump centered at $35 \pm 2.3\,M_\odot$ to test/reproduce LVK observations (see Paper I). We set $\gamma = 2$ by default, and we also test a heavier IMF of $\gamma = 1$. Since mass segregation can strongly affect the BH IMF, we also test an IMF characterized by strong mass segregation by setting $P(M) \propto M^{+2.5}$ (B. Rom & R. Sari 2025) spanning [10, 40] $M_\odot$, also tested by V. Delfavero et al. (2025b). Finally, we test a BH IMF constrained to the mass pileup observed by LVK, drawn from a Gaussian distribution about $34 \pm 4\,M_\odot$ (S. Kishore Roy et al. 2025). IMFs dominated by mass segregation have heavier BHs on average, with larger average Hill radii ($R_{\rm H}$) and faster migration rates. We therefore expect the most efficient chain of hierarchical mergers to result from the mass-segregated BH IMFs.

### 3.7. Initial Spin Distribution Models

The natal spin distribution of BHs is not well understood. Here we follow the default model in Papers I and II, where spin magnitudes are drawn from a Gaussian centered on $a = 0$, with variance $\sigma_a$. We make use of the semianalytical precession model outlined in D. Gerosa & M. Kesden (2016) to calculate more accurately ($a$, $v_{\rm kick}$) of remnant BHs. We change $\sigma_a$ to investigate the impact on expected jet and shock luminosities.

## 4. Results

In this section we present the bolometric jet and shock luminosities produced in a fiducial McFACTS v.0.4.0 simulation and their rates of likely detectability, and we examine how variations in the underlying physical parameters impact these distributions.

For convenience, if the number density of AGNs is approximately $n_{\rm AGN} \sim 10^{-3}\,{\rm Mpc}^{-3}$–$10^6\,{\rm Gpc}^{-3}$ (B. McKernan et al. 2018), and if the average AGN disk lasts ~1 Myr, a useful merger rate unit is $\mathcal{R}_{\rm GW} = 1$ merger/AGN/Myr, which corresponds to $\mathcal{R}_{\rm GW} = 1$ Gpc$^{-3}$ yr$^{-1}$, or ~2%–4% of the observed LVK rate (The LIGO Scientific Collaboration et al. 2026b).

Various results of the default/fiducial McFACTS v.0.4.0 simulation, using the parameters in Table 1, are presented in Figures 2–4, with variations of the default parameters in later plots. We follow Papers I–III by assigning marker colors in all figures according to merger generation, where gold points correspond to first-generation (1g–1g) mergers, purple points involve at least one second-generation BH (2g), and red points correspond to mergers involving at least one third- or higher-generation BH ($\geqslant$3g). The horizontal dashed–dotted line corresponds to $L_{\rm AGN} = 1.26 \times 10^{45}\,{\rm erg\,s^{-1}}$ in plots showing the jet luminosity.

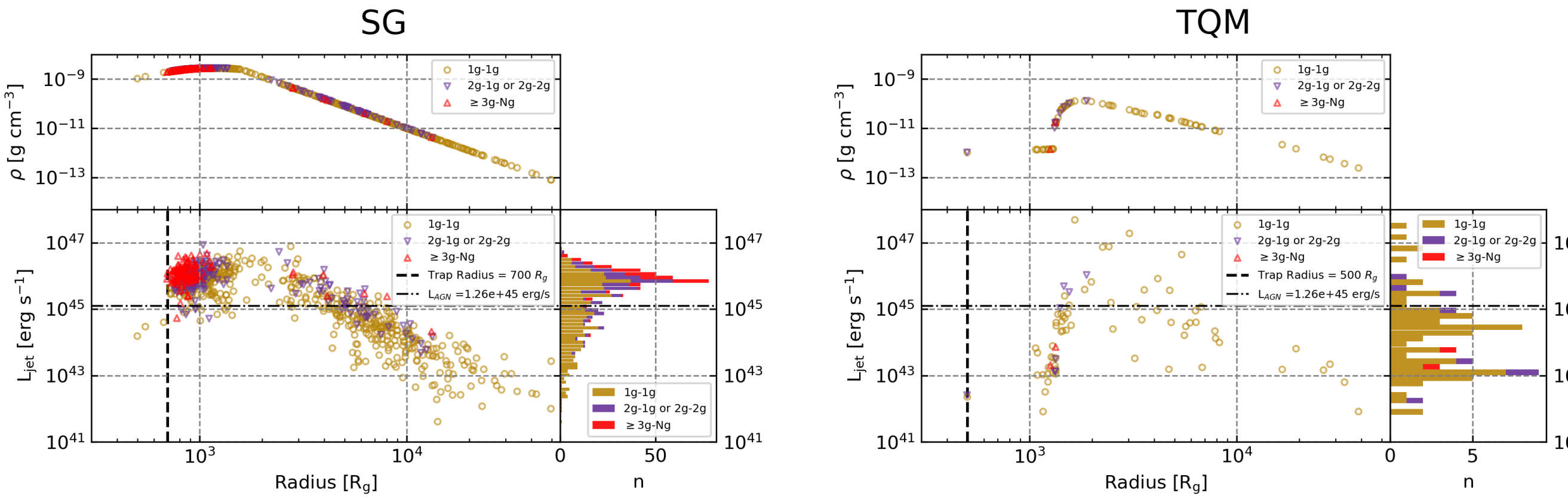

**Figure 2.** Jet luminosity distribution and disk density as a function of BH semimajor axis. Left set of panels: SG disk model. The corresponding merger rate is $\mathcal{R}_{\rm GW} \sim 11$ Gpc$^{-3}$ yr$^{-1}$. Right set of panels: TQM disk model. The corresponding merger rate is $\mathcal{R}_{\rm GW} \sim 1$ Gpc$^{-3}$ yr$^{-1}$. Upper panels: disk model radial density profile. Central panels: bolometric jet luminosity (erg s$^{-1}$) vs. radius in disk ($R_g$) calculated for each merger in McFACTS. Gold represents 1g–1g mergers, purple represents 2g–$m$g ($m \leqslant 2$), and red represents 3g–$n$g ($n \in \mathbb{Z}^+$), where 1g is a first-generation BH (never experienced a merger), 2g is a second-generation BH (previously experienced one merger), etc. Right panels: number of BBH mergers per generation of BHs as a function of jet luminosity. These distributions represent mergers in 100 AGNs with $M_{\rm SMBH} = 10^8\ M_\odot$ over 0.7 Myr. Galaxy num = 100, and seed = 3456789018.

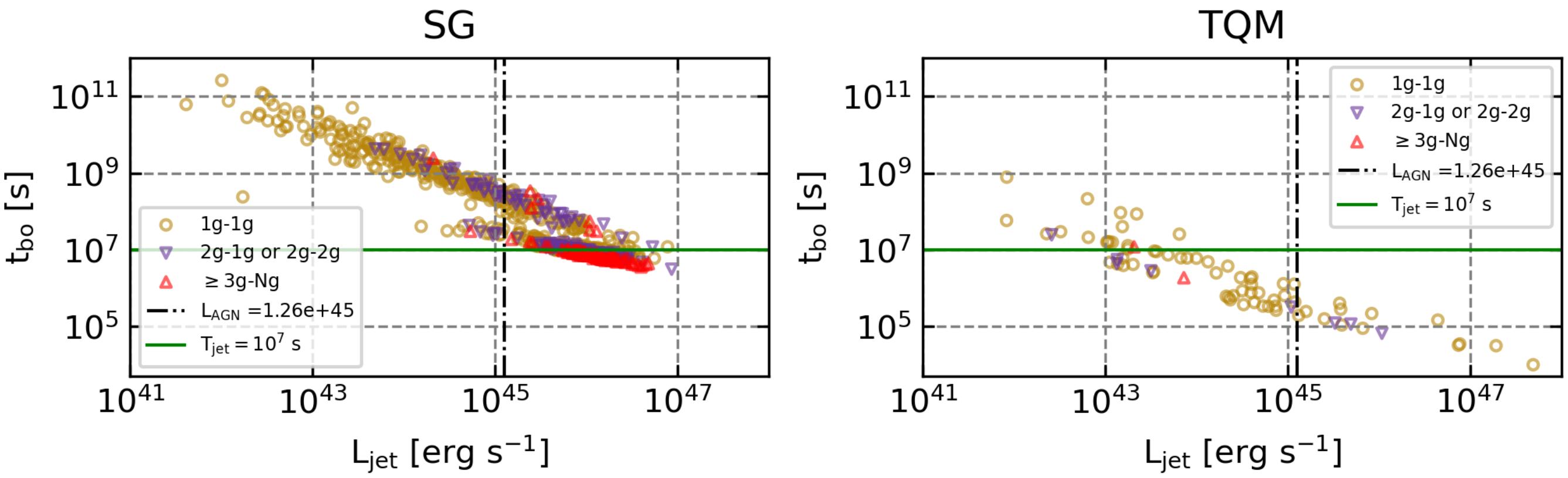

**Figure 3.** Jet breakout time as a function of jet luminosity. Left: SG disk model. Right: TQM disk model. Our fiducial jet lifetime, $T_{\rm jet} = 10^7$ s, is indicated with a solid green line. If the breakout time is shorter than the jet lifetime, the jet is expected to break out of the disk and its emission can be potentially observable. If the breakout time is longer than the jet lifetime, then the jet emission escapes the disk on the photon diffusion time, and it is practically undetectable.

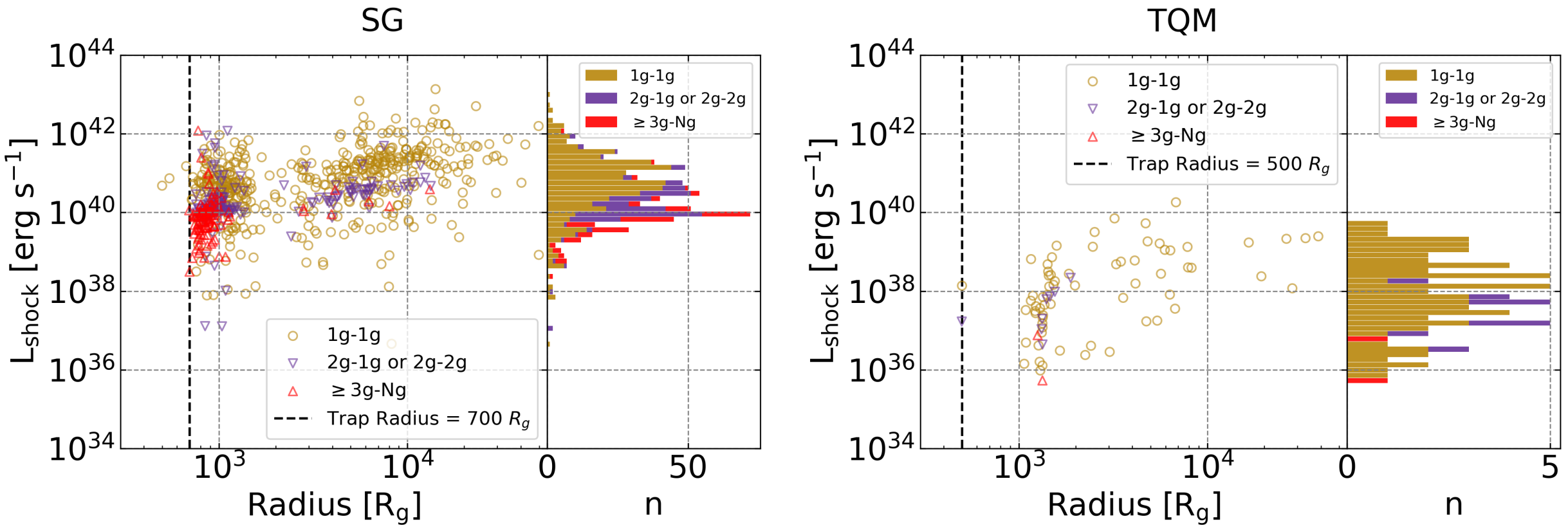

**Figure 4.** Shock luminosity distribution as a function of BH semimajor axis. Left: SG disk model. Right: TQM disk model. Left panels: Bolometric shock luminosity (erg s$^{-1}$) as a function of disk radius ($R_g$) for each merger in McFACTS. Right panels: Number of BBH mergers per BH generation as a function of shock luminosity.

### *4.1. Disk Model*

Figure 2 shows $L_{\rm jet}$ as a function of merger radial location for both disk models.

Two features are apparent in the left set of panels of Figure 2. The clump of (mostly red) points around $10^3 R_g$ results from the migration swamp in the SG disk model, where migration speed decreases substantially before reaching the trap (vertical dashed line, $R = 700R_g$) (see Paper I). The broader clump at larger radii consists of mergers in the bulk disk. Note that the influence of the migration swamp encourages efficient hierarchical merging (which should be a source of high-spin BH and potential jet formation). The top panel of Figure 2 shows $L_{\rm jet}$ as a function of disk density and merger generation. The right set of panels of Figure 2 shows a two-peaked distribution of $L_{\rm jet}$, corresponding mostly to the difference between mergers at the migration swamp and those in the bulk disk. The highest-generation mergers (red points) typically have the highest masses and highest spins of the total population, yielding the largest $L_{\rm jet}$ around the migration swamp and permitting kinetic jet luminosities potentially orders of magnitude greater than that of the host AGN.

In the right set of panels of Figure 2, a sharp transition in the TQM disk model density profile produces a migration swamp near $10^3 R_g$. Although this enables a small number of hierarchical mergers, the frequency of $\geqslant$2g mergers is strongly suppressed owing to the comparatively low density of the TQM disk. This lower density also increases the circularization timescale relative to the default disk lifetime, resulting in an overall merger rate substantially lower than that obtained for the SG model.

Figure 3 shows breakout time ($t_{\rm bo}$) as a function of $L_{\rm jet}$. The jet is able to emerge from the disk (and is therefore potentially detectable) if $t_{\rm bo} < T_{\rm jet}$ (horizontal green line). In the SG disk (Figure 3, left), most low-generation BBH mergers, as well as most mergers in the bulk disk population, have $t_{\rm bo} > T_{\rm jet}$. In these cases, the injected energy emerges on the photon diffusion timescale, indistinguishable from ordinary AGN variability. In the TQM disk (Figure 3, right), the corresponding region has an aspect ratio roughly an order of magnitude smaller than in the SG model, resulting in significantly shorter breakout timescales for the bulk merger population.

Figure 4 shows the distribution of shock luminosities $L_{\rm shock}$ for our fiducial model. For the SG disk model (Figure 4, left panel), only a small number of mergers produce $L_{\rm shock}$ values potentially comparable to Seyfert AGN luminosities ($>10^{42}$ erg s$^{-1}$), and none approach $L_{\rm AGN}$ in our fiducial disk model. In the TQM disk model, no mergers yield $L_{\rm shock}$ comparable to the assumed $L_{\rm AGN}$. Nevertheless, these events are expected to contribute to stochastic heating and turbulence within AGN disks.

To assess how variations in the initial population affect potentially observable flares, we next investigate how changes in `McFACTS` input parameters modify the predicted luminosity distribution. Since, as stated earlier, shock luminosities are found to be subdominant to the luminosity of the host AGN, for the remainder of our analysis we simply focus on how parameter variations impact the $L_{\rm jet}$ distribution.

### *4.2. Dependence on the Disk Lifetime*

Here we explore how variations in the accretion disk lifetime ($\tau_{\rm AGN}$) influence the merger population and therefore the possible jet luminosities produced by embedded BBH remnants.

Figure 5 shows the effect of varying the disk lifetime from $\tau_{\rm AGN} = 0.25$ to 1 Myr, with the default $\tau_{\rm AGN} = 0.7$ Myr in the middle panels. The top row displays the results for the SG disk, where, from left to right, the corresponding runs are `sg_tagn_025Myr`, `sg_default`, and `sg_tagn_1Myr`. For a short-lived 0.25 Myr disk (Figure 5(a)), the number of higher-generation (and therefore high-spin) mergers is significantly reduced; these systems typically produce the brightest jet luminosities, and therefore only a small number of mergers generate luminous jets. In contrast, for the slightly longer-lived disk (top right panel), we find an increase in luminous mergers. The extended lifetime allows hierarchical merging to proceed for a longer period—particularly within migration traps/swamps, where the SG disk reaches densities of order of $10^{-9}$ g cm$^{-3}$ (see the upper panel in the left set of panels of Figure 2)—thereby promoting the production of bright jets. Prolonged hierarchical merging yields many more higher-generation remnants with the high mass and high spin required to power significant jet luminosities.

In the bottom row of Figure 5 we display the results for the TQM model, where, from left to right, the corresponding runs are `tqm_tagn_025Myr`, `tqm_default`, and `tqm_tagn_1Myr`. Lower surface density leads to slower migration rates and thus an overall smaller merger rate across choices of disk lifetime. In a short-lived 0.25 Myr disk (Figure 5(d)), no binaries have time to form and merge. Whereas binaries are able to form and merge $\leqslant$0.3 Myr in the SG disk, mergers begin around $\sim$0.5 Myr in the TQM model (Figure 5(e)). A prolonged disk lifetime (Figure 5(f)) allows for the formation and merger of more BBHs for an enhanced rate of 1g mergers, as well as the persistence of a hierarchical merger chain, creating more $\geqslant$2g mergers.

Regardless of disk model choice, changing the disk lifetime affects both the total number of mergers and the duration over which hierarchical merger chains can proceed. Thus, as the merger rate increases, the number of mergers capable of producing potentially bright EM counterparts grows nonlinearly with disk lifetime.

### *4.3. Dependence on the Disk Size*

Next, we examine how varying the outer-disk radius affects the predicted bolometric jet luminosities. We adjust $R_{\rm outer}$ (`disk_radius_outer` in `McFACTS`) to a smaller value, $3 \times 10^4 R_g$, and to a larger value, $7 \times 10^4 R_g$. Changing the disk extent alters the number of initial BHs embedded in the disk and consequently modifies the resulting hierarchical merger chain.

Figure 6 shows the results for different outer-disk radii. The corresponding runs from left to right are `sg_r3e4`, `sg_default`, and `sg_r5e4`. In the first panel, corresponding to an outer-disk radius of $R_{\rm outer} = 3 \times 10^4 R_g$, there are fewer BBHs in the outer disk than in the default run. A smaller disk radius decreases the cross section between the NSC and the accretion disk, limiting the number of BHs that can be captured in the outer disk. However, the disk capture rate (in the inner disk) remains unchanged. As a result, the overall merger rate is lower; however, migration still drives efficient hierarchical mergers at the migration trap and swamp. With the smaller disk, a larger fraction of mergers occur at the trap, and some mergers now take place in the inner disk. Because hierarchical merging at the migration swamp and trap is not strongly suppressed, the number of potentially observable counterparts in these regions remains largely unchanged. While outer-disk mergers are reduced in this model, these events are less likely to break out.

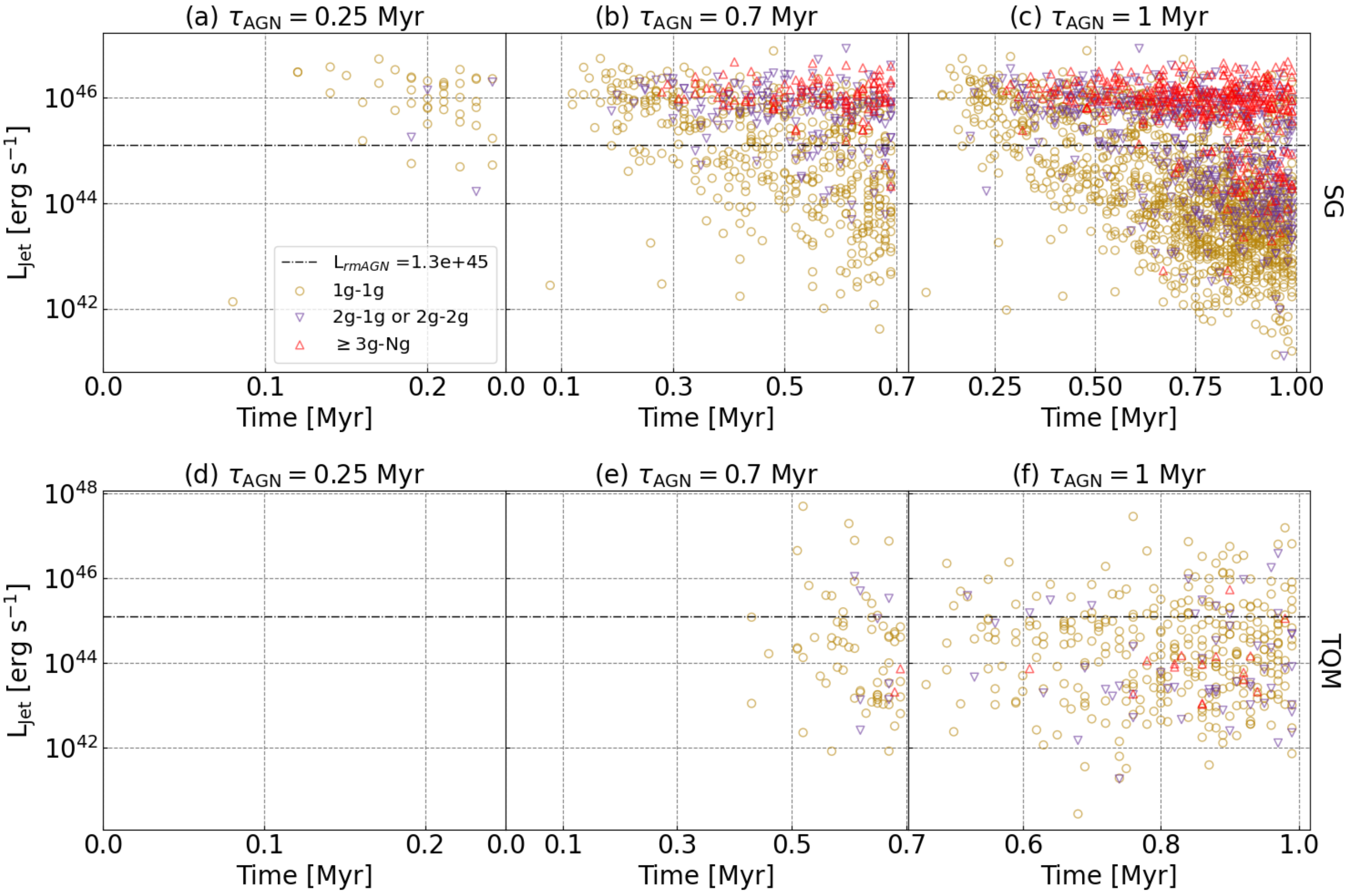

**Figure 5.** Effects of the disk lifetime. Bolometric jet luminosity (erg s$^{-1}$) is shown as a function of merger time (Myr), for disks with lifetimes $\tau_{\rm AGN} = 0.25$, 0.7, and 1 Myr (left to right). Top row: SG disk model. The corresponding merger rates are $\mathcal{R}_{\rm GW} \sim 1\ {\rm Gpc^{-3}\ yr^{-1}}$, $\mathcal{R}_{\rm GW} \sim 11\ {\rm Gpc^{-3}\ yr^{-1}}$, and $\mathcal{R}_{\rm GW} \sim 23\ {\rm Gpc^{-3}\ yr^{-1}}$. Bottom row: TQM disk model. The corresponding merger rates are $\mathcal{R}_{\rm GW} \sim 0\ {\rm Gpc^{-3}\ yr^{-1}}$, $\mathcal{R}_{\rm GW} \sim 1\ {\rm Gpc^{-3}\ yr^{-1}}$, and $\mathcal{R}_{\rm GW} \sim 4\ {\rm Gpc^{-3}\ yr^{-1}}$.

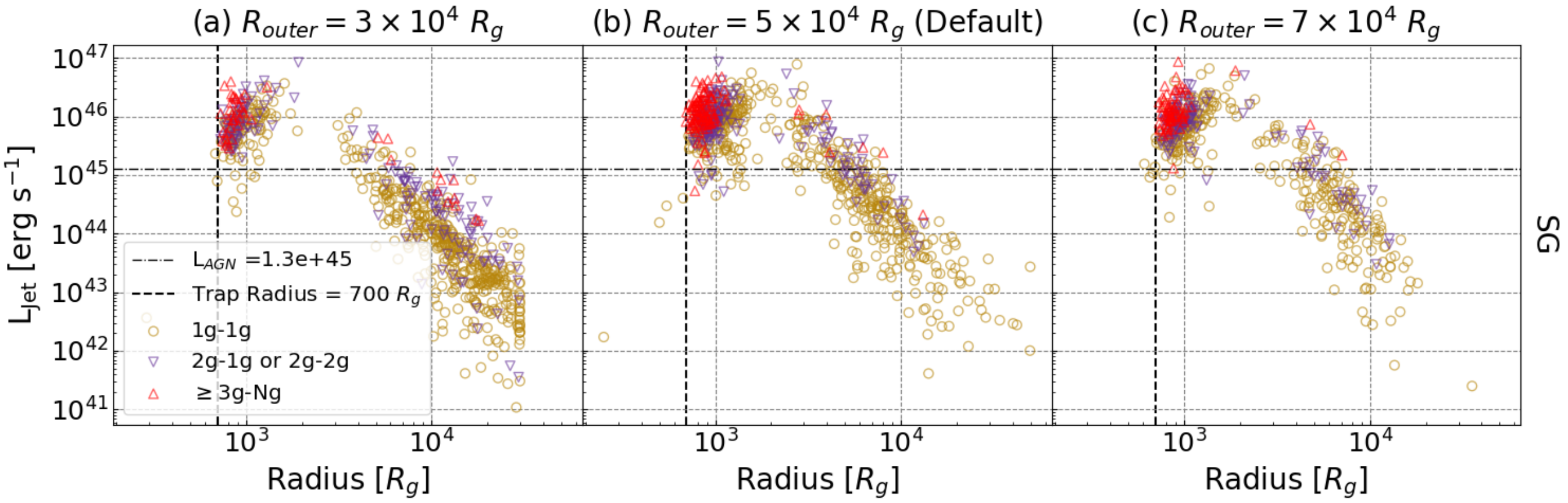

**Figure 6.** Effect of the radial extent of the accretion disk. Bolometric jet luminosity (erg s$^{-1}$) is shown as a function of radial distance ($R_g$), for disks of different sizes: $R_g = 3 \times 10^4\ R_g$, $5 \times 10^4\ R_g$, $7 \times 10^4\ R_g$ (from left to right). The corresponding merger rates are $\mathcal{R}_{\rm GW} \sim 10\ {\rm Gpc^{-3}\ yr^{-1}}$, $\mathcal{R}_{\rm GW} \sim 11\ {\rm Gpc^{-3}\ yr^{-1}}$, and $\mathcal{R}_{\rm GW} \sim 7\ {\rm Gpc^{-3}\ yr^{-1}}$.

With an extended disk (see Figure 6(c)) of radius $7 \times 10^4\ R_g$, despite the larger number of initial BHs, binaries have a smaller chance to form owing to the relatively larger average initial separations between BHs in the disk. As a result, the number of binaries formed is depleted and hierarchical merging is discouraged, in turn leading to a smaller number of high-spin remnants. Thus, while the number of possible EM counterparts brighter than $L_{\rm AGN}$ is slightly depleted compared to the default run (see Figures 6(b) and (c)), the number of possible counterparts is comparable between runs.

While we do not offer direct analysis of the TQM disk model because of specific disk physics sensitive to the choice of $R_{\rm outer}$, we suggest that these results should be analogous between choices of outer radius in TQM as well, as the cross section of NSC objects coincident with the disk is still expected to scale with outer radius.

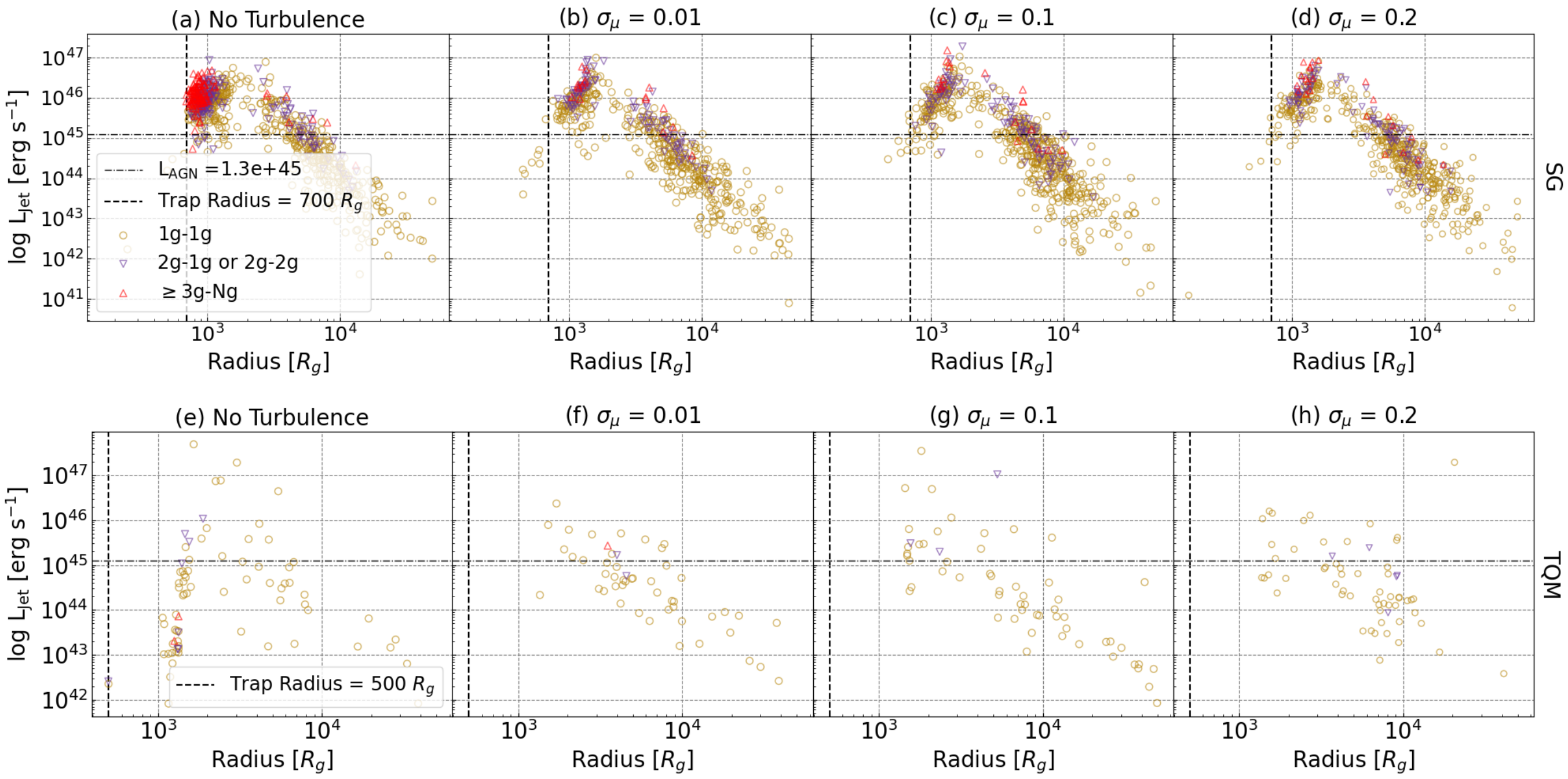

**Figure 7.** Effects of disk turbulence strength. Bolometric jet luminosity (erg s$^{-1}$) is shown as a function of remnant semimajor axis ($R_g$) measured from the central SMBH, for models in which the migration prescription includes a stochastic component to mimic disk turbulence. The width of the Gaussian governing the turbulent migration term is varied between panels. The leftmost panels show the models without turbulence, while the subsequent panels correspond to progressively more turbulent disks, with $\sigma_{\rm migration} = 0.01, 0.1, 0.2$ (left to right). Top row: SG disk model. The corresponding merger rates are all $\mathcal{R}_{\rm GW} \sim 10$ Gpc$^{-3}$ yr$^{-1}$. Bottom row: TQM disk model. The corresponding merger rates are all $\mathcal{R}_{\rm GW} \sim 1$ Gpc$^{-3}$ yr$^{-1}$.

### *4.4. Effect of Disk Turbulence*

Figure 7 compares our fiducial run, which does not include a stochastic term, to three runs where we sequentially increase the amount of "turbulence" per run. The top row of Figure 7 displays the results for the SG model. The corresponding runs from left to right are `sg_turb01` and `sg_default`, and the bottom row corresponds to runs `sg_turb1` and `sg_turb2`. In the default SG run (Figure 7(a)), the migration trap at $700R_g$ and the migration swamp near $10^3R_g$ create pileup features as embedded BHs stall in the disk, form binaries, and merge. These BHs promote hierarchical mergers and thus faster-spinning remnants. This feature is smeared by the introduction of and increase in turbulence across subsequent runs.

Figure 7 displays the results of varying disk turbulence for the TQM model. The corresponding runs from left to right are `tqm_turb01`, `tqm_default`, while the bottom row corresponds to runs `tqm_turb1` and `tqm_turb2`. The sharp hump-like feature around $10^3R_g$ corresponding to the steep change in density around the region is clearly smeared out by increasing turbulence in the disk, looking across Figures 7(e)–(h).

Broadly, Figure 7 shows that the introduction of stochastic turbulence on top of migration smears out and shifts radially outward the clump of hierarchical mergers at the migration swamp (around $\sim 10^3R_g$). Likewise, the hierarchical mergers at the migration trap are suppressed by turbulence and also smeared out. Generally, as turbulence is increased, there is a slight increase in the variance of $L_{\rm jet}$ as a function of a given merger radius.

### *4.5. Dependence on the Initial Mass Function*

Next, we investigate how varying the initial BH mass function affects the resulting bolometric jet luminosities. We modify the IMF to follow $\gamma \propto M^{-1}$, $M^{+2.5}$ by adjusting the parameter `nsc_imf_powerlaw_index`.

The results for varied BH IMF are compared in Figure 8. The top row of Figure 8 shows the SG results; the corresponding runs from left to right are `sg_default`, `sg_g1`, `sg_g25`, and `sg_gpeak`. The bottom row of Figure 8 shows the TQM results; the corresponding runs from left to right are `tqm_default`, `tqm_g1`, `tqm_g25`, and `tqm_gpeak` The steep, negatively sloped IMF in our default model (Figures 8(a) and (e)), $\gamma \propto M^{-2}$, biases the initial BH masses toward lower values. Lower initial masses lengthen the timescale required to build high-mass, high-spin remnants through hierarchical mergers and result in a remnant population with a lower average mass compared to the other IMFs we explore.

With a flatter IMF of $M^{-1}$ (Figures 8(b) and (f)), higher-mass BHs can be initially assigned to objects embedded in the disk. This shifts the initial BH masses toward higher values, producing even more massive remnants as these BHs merge. Because `McFACTS` assumes a binary to form when the Hill radii of two BHs overlap, and because the Hill radius scales with mass, the higher masses lead to more frequent mergers. This increased merger rate efficiently builds up BH mass and spin, thus creating a population potentially capable of driving bright jets.

The next IMF we explore has the functional form $M^{+2.5}$. Figures 8(c) and (g) show that a steep, positively sloped IMF of $M^{+2.5}$ biases the initial BH population toward substantially larger masses. The resulting larger Hill radii enhance the probability of binary formation, thereby promoting hierarchical mergers. This efficiently pumps up the masses and spins of the embedded population, which are key for producing bright jets, especially in the SG case (Figure 8(c)). Under this steep IMF, we find that 1g mergers occur at significantly higher masses, increasing the number of potentially bright counterparts from 1g mergers (if only evaluating $L_{jet}$ against $L_{AGN}$). Distinct features emerge in the 2g and $\geqslant$3g mergers, reflecting the buildup of mass through successive mergers at the migration swamp in both disk models. The lower-mass component originates primarily in the outer disk in SG and across the disk in TQM.

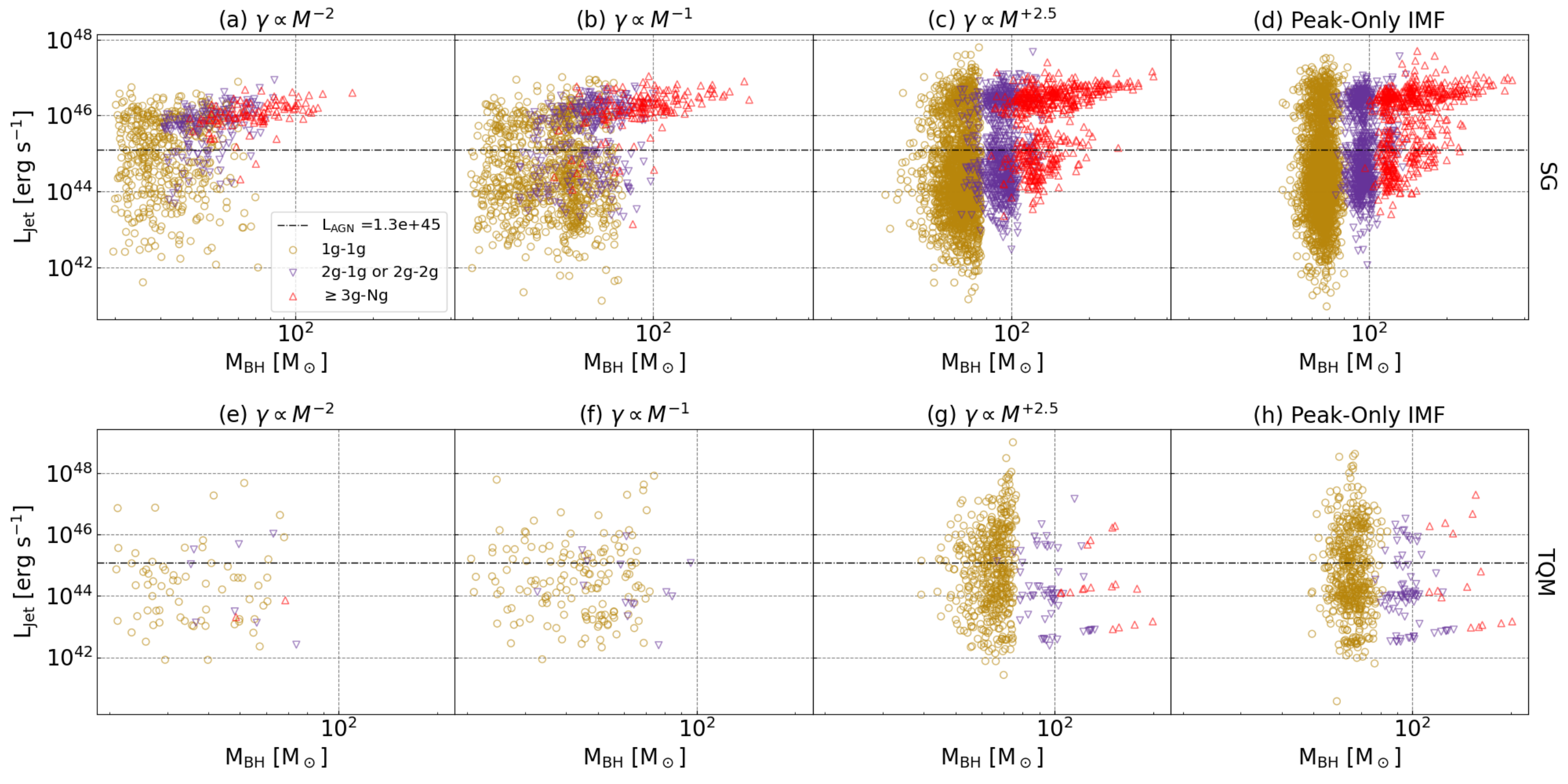

**Figure 8.** Dependence of remnant mass on the NSC BH IMF. Jet luminosity (erg s$^{-1}$) is shown as a function of remnant mass. From left to right, the IMF is varied according to $\gamma \propto M^{-2}$, $M^{-1}$, $M^{+2.5}$ and a peak-only power law with $M^{-2}$. Top row: SG disk model. The corresponding merger rates are $\mathcal{R}_{GW} \sim 11$ Gpc$^{-3}$ yr$^{-1}$, $\mathcal{R}_{GW} \sim 21$ Gpc$^{-3}$ yr$^{-1}$, $\mathcal{R}_{GW} \sim 51$ Gpc$^{-3}$ yr$^{-1}$, and $\mathcal{R}_{GW} \sim 54$ Gpc$^{-3}$ yr$^{-1}$. Bottom row: TQM disk model. The corresponding merger rates are $\mathcal{R}_{GW} \sim 1$ Gpc$^{-3}$ yr$^{-1}$, $\mathcal{R}_{GW} \sim 2$ Gpc$^{-3}$ yr$^{-1}$, $\mathcal{R}_{GW} \sim 8$ Gpc$^{-3}$ yr$^{-1}$, and $\mathcal{R}_{GW} \sim 9$ Gpc$^{-3}$ yr$^{-1}$.

Lastly, we consider an IMF that restricts the initial masses to the 34 $\pm$ 4 $M_\odot$ peak observed by LVK. Figures 8(d) and (h) show a top-heavy, peak-only IMF, where $\sim$35 + 35 $M_\odot$ mergers dominate the first generation of BHs. As in the $M^{+2.5}$ case, features in the 2g and $\geqslant$3g mergers result from the buildup of mass through successive mergers at the migration swamp.

Overall, the IMF strongly regulates the efficiency with which BHs grow through mergers, and as the remnant masses increase, the jet luminosities rise proportionally.

### *4.6. Dependence on the Initial Spin Distribution*

Finally, we explore how altering the width of the Gaussian distribution from which the initial spins are drawn impacts the jet luminosity. By modifying the parameter `nsc_bh_spin_dist_sigma` in `McFACTS`, we adjust the spread of the initial spin distribution and therefore the dispersion of 1g merger spins.

Figure 9 shows the results for spin distribution variations. The top row displays the results for the SG disk model; the corresponding runs from left to right are `sg_spin_dist_sigma_02`, `sg_default`, and `sg_spin_dist_sigma_2`. The bottom row displays the corresponding results for the TQM disk; the corresponding runs from left to right are `tqm_spin_dist_sigma_02`, `tqm_default`, and `tqm_spin_dist_sigma_2`. As the Gaussian is broadened (Figures 9(a)–(c) and (d)–(f)), with width increasing left to right), the spin values of 1g mergers diffuse. The broadest distribution ($\sigma_a = 0.2$) shifts the 1g merger spins to lower values, due to the initial population containing a larger spread of spins. Because the jet luminosity scales as $a^2$, this produces a corresponding reduction in jet power.

In contrast, a narrower initial distribution, $\sigma_a = 0.02$, yields a slightly more concentrated set of initial spins and preserves features associated with migration-trap clustering in the disk. Across all spin distributions, the impact on higher-generation spins is comparatively modest, as these remnants acquire their spins primarily through hierarchical mergers, with the dominant contribution arising from the orbital angular momentum of the binaries that formed them. In most cases, the binary orbital angular momentum is aligned with the angular momentum of the disk. Additionally, these higher-generation BHs have been further spun up by accretion in the disk.

Changing the initial spin distribution in the disk has minimal impact on the overall merger rate or on the formation of higher-generation remnants. Consequently, the jet luminosities from the brightest $\geqslant$3g merger products remain essentially unchanged in the SG disk model, whereas bright jet production is diminished in the TQM disk because of a greater chance for asymmetric spins, proportionally decreasing the 1g spins and consequently diminishing the remnant jet power across TQM remnant generations.

## 5. Discussion: Correlating GW and EM Counterparts to BBH Mergers

At the population level, we can identify the types of mergers most likely to produce observable EM counterparts. In our model, high-mass, high-spin BBH remnants—particularly 2g and ⩾3g mergers—are the dominant sources of significant jet luminosities.

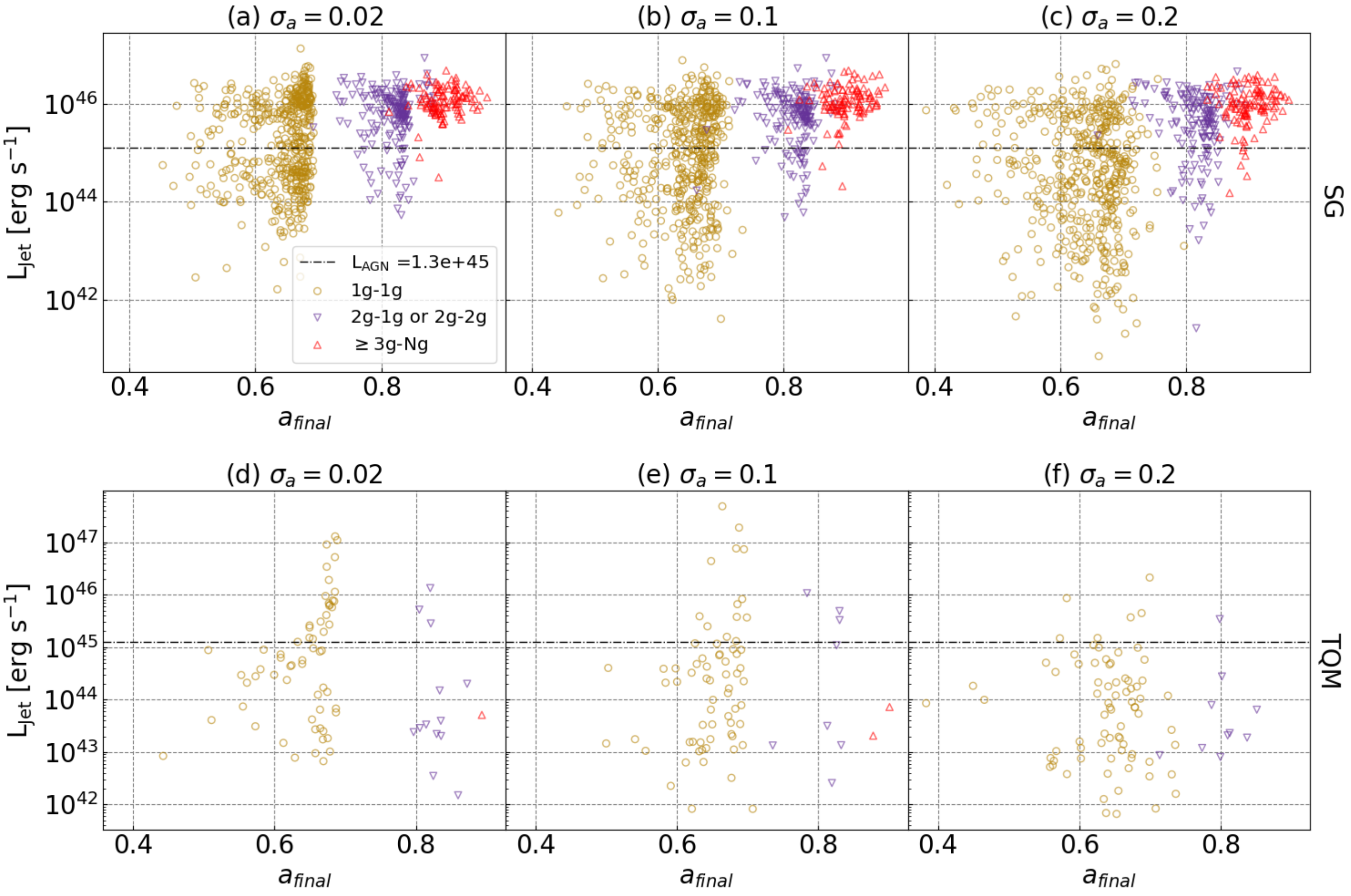

**Figure 9.** Dependence on the initial spin distribution. Jet luminosity (erg s$^{-1}$) as a function of BH spin. From left to right, the width of the Gaussian used to sample initial BH spins is varied to $\sigma_x = 0.02$, 1, 0.2. Top row: SG disk model. The corresponding merger rates are all $\mathcal{R}_{GW} \sim 11$ Gpc$^{-3}$ yr$^{-1}$. Bottom row: TQM disk model. The corresponding merger rates are all $\mathcal{R}_{GW} \sim 1$ Gpc$^{-3}$ yr$^{-1}$.

Throughout this work, we adopt the criterion $L_{jet} > L_{AGN}$ as a *fiducial threshold* for identifying potentially observable jet-powered counterparts. We emphasize, however, that this condition should not be interpreted as a rigorous detectability criterion. In practice, the observability of a jet depends on a variety of factors that are not specified in our model, including the spectral energy distributions of both the jet and the AGN, observing band and instrument sensitivity, source distance, extinction, variability of the AGN background, and measurement uncertainties. A jet with $L_{jet} < L_{AGN}$ may still be detectable if its emission is concentrated in a wavelength range where the AGN is relatively faint, while a jet with $L_{jet} > L_{AGN}$ may remain difficult to identify if its emission is distributed across poorly observed bands or is obscured by AGN variability. Consequently, our adopted threshold serves only as a convenient reference point for comparing models and estimating the relative likelihood of EM counterpart production, rather than as a definitive prediction of observational detectability.

The top panel of Figure 10 shows the fraction of mergers with observable flares per Grace-DB[14] chirp mass $\mathcal{M}$ bin, as well as the kinetic jet luminosity, remnant spin, and jet breakout time for all mergers in our fiducial model, as a function of $\mathcal{M}$ in subsequent panels. Figure 10 represents the full sample of simulated mergers for both the SG (left) and TQM (right) default runs, with the subset of merger remnants with observable EM counterparts color-coded by merger generation, while gray symbols represent muffled, unobservable points. Colored symbols (gold, purple, and red for sequential merger generations; see discussion in Section 3) indicate the subset of merger remnants whose jets both break out before the jet turns off ($t_{bo} < T_{jet}$) and exceed the AGN luminosity ($L_{jet} > L_{AGN}$). Notably, in the SG run all BHs with $t_{bo} < T_{jet}$ also have jet luminosities $L_{jet} > L_{AGN}$, while the TQM model produces a small number of mergers where $t_{bo} < T_{jet}$ but $L_{jet} < L_{AGN}$ (see Figures 2 and 3). We include a point with the GW observable from GW190521, as well as the EM observables from the most well-studied optical counterpart candidate ZTF19abanrhr ($t_{bo}$, where we assume that the breakout time is the rest-frame photon diffusion time; see M. J. Graham et al. 2020, for reference) for comparison with our model's predictions.

We present two model realizations for discussion: the default model and the peak-only IMF model. Both models show that the breakout time is fully consistent with our $t_{bo}$ model (see Figures 10 and 11). The GW190521 + ZTFabanrhr point falls within $2\sigma$ of the spin and jet luminosity distributions of our default model.

Figures 10 and 11 suggest that GW190521 could be the product of a 2g or 3g merger originating in an SG-like disk. In the peak-only scenario (Figure 11, left), we would expect a slightly higher spin value than reported, but the spin is in agreement at the 2$\sigma$ level. An assumption of larger spin has little impact on the inferred chirp mass (C. Cutler & É. E. Flanagan 1994). Additionally, because $L_{\rm jet}$ is lower than expected by our model (second panels of Figures 10 and 11), we propose the the jet could have been observed off-axis, which would explain this deficit.

[14] https://gracedb.ligo.org/

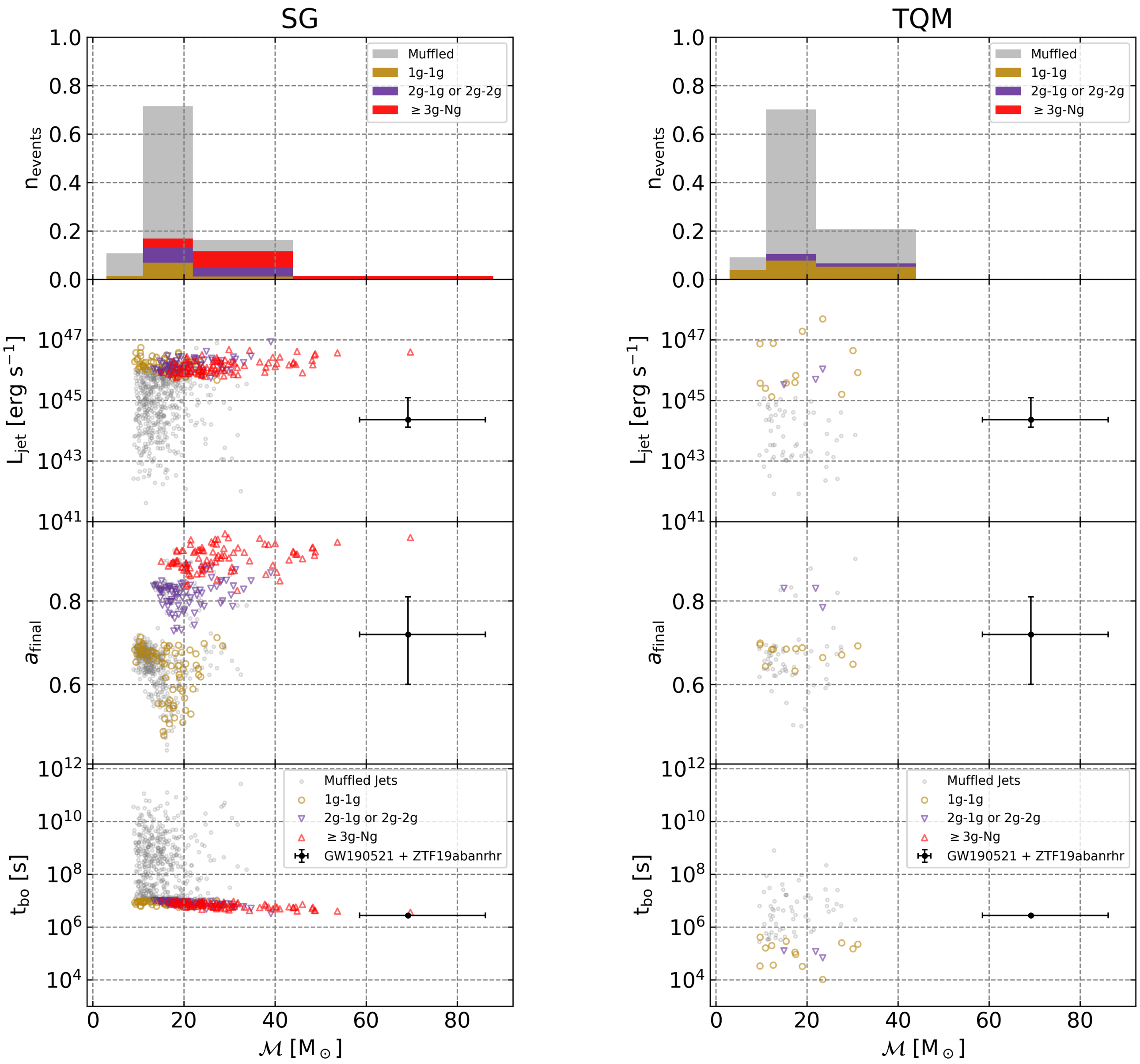

**Figure 10.** Jet luminosity, remnant spin, breakout time, and fraction of EM bright/dark mergers as a function of chirp mass for the fiducial model. Gray symbols represent muffled jets and are unobservable. Colored symbols indicate mergers with $L_{\rm jet} > L_{\rm AGN}$ and $t_{\rm bo} < T_{\rm jet}$ for our fiducial $T_{\rm jet} = 10^7$ s. Top panel: colored histograms indicate the relative fraction of mergers expected to have observable jet emission, for each generation of merger. The fraction of events with choked jets is represented by the gray histogram. Histogram bins represent Grace-DB chirp mass bins, $\mathcal{M} = [3.0–11.0, 11.0–22.0, 22.0–44.0, 44.0–88.0]$ $M_\odot$. Left column: SG disk model. The corresponding percentages of bright mergers per chirp mass bin are 14%, 24%, 71%, 100%. Right column: TQM disk model. The corresponding percentages of bright mergers per chirp mass bin are 43%, 15%, 31%, 0%.

Accounting for geometrical observing constraints, $\sim$1/2 of these EM flares will be obscured in Type II AGNs, thus reducing the quoted probabilities. Another $\sim$1/2 of these observable flares will emerge on the side of the disk opposite to the observer if imparted with a kick opposite to the observer, which could conservatively limit the number of practically observable events by 3/4 (1/2 from obscured AGNs, at least 1/2 from viewing angle effects).

Our results are broadly consistent with current observational follow-up studies. In particular, T. Cabrera et al. (2026) revisited associated EM flares in M. J. Graham et al. (2023) to constrain the fraction of optical counterparts detectable by ZTF to $\sim$3%, implying that as many as $\sim$40% of BBH mergers may occur within AGN disks. If this scenario is correct, and if the bulk of the jet emission emerges in the optical band, then the baseline likelihoods for our fiducial SG (TQM) models (Table 2) imply that observable counterparts would accompany $\sim$3% (2%) of all BBH mergers in a Universe where AGN disks are predominantly SG-like (TQM-like). If, on the other hand, we use GWTC-5 and we associate the high-mass, high-spin population with AGN disks, then more likely $f_{\rm BBH,AGN} \sim \mathcal{O}(1-10)\%$, which could suggest that ∼1% of all BBH mergers could be accompanied by observable EM counterparts. Dedicated statistical study is warranted to constrain this association in the most recent observing runs.

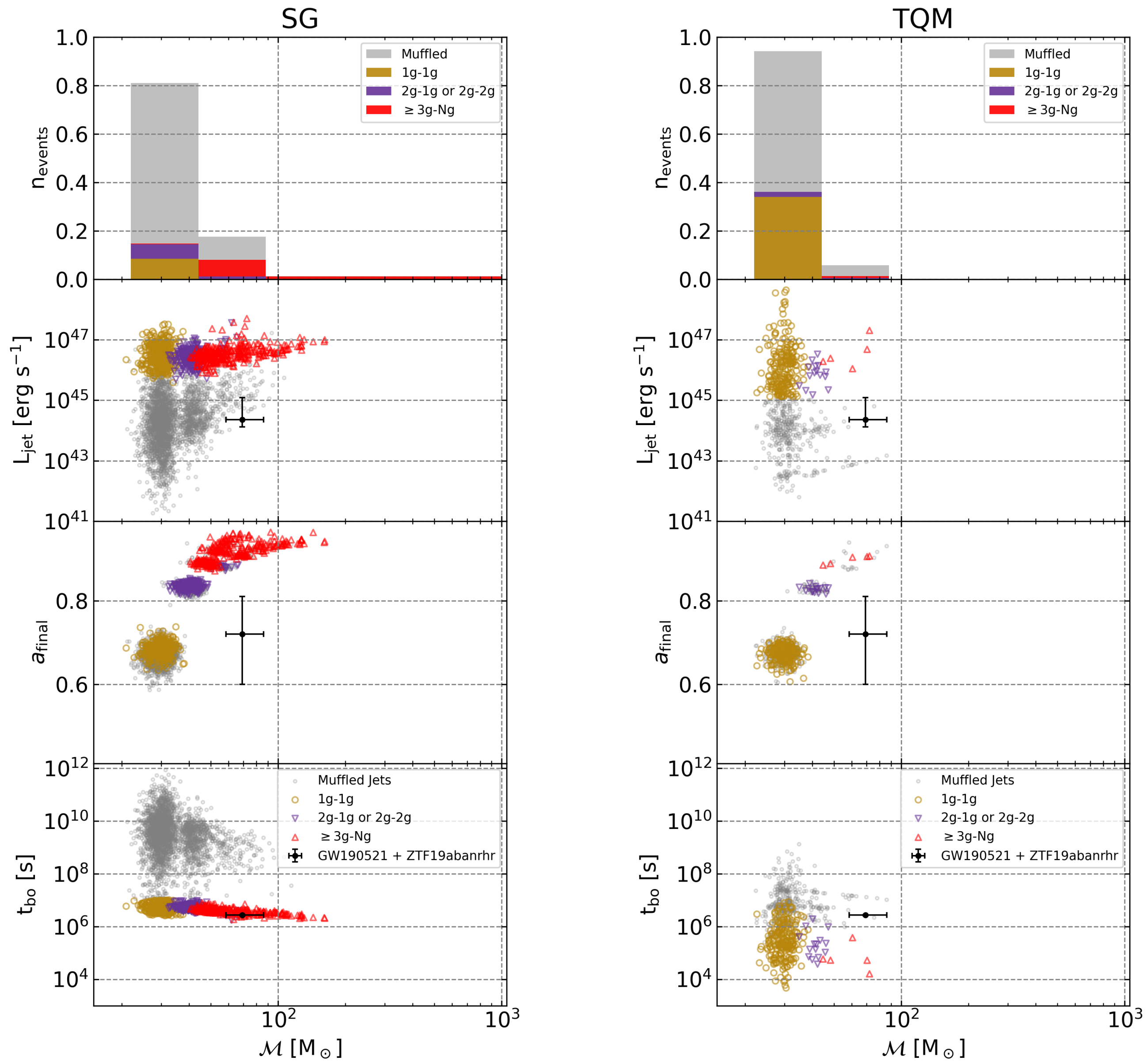

**Figure 11.** Jet luminosity, remnant spin, breakout time, and fraction of EM bright/dark mergers as a function of chirp mass for the peak-only IMF. Histogram bins represent Grace-DB chirp mass bins, $\mathcal{M}$ = [3.0–11.0, 11.0–22.0, 22.0–44.0, 44.0–88.0, 88.0–1000.0] $M_\odot$. Left column: SG disk model. The corresponding percentages of bright mergers per chirp mass bin are 0%, 33%, 18%, 46%, 87%. Right column: TQM disk model. The corresponding percentages of bright mergers per chirp mass bin are 0%, 0%, 38%, 23%, 0%.

Our results suggest a correlation between disk model and chirp mass, evident in Figure 10. For Grace-DB-like chirp mass bins of $\mathcal{M}$ = [3.0–11.0, 11.0–22.0, 22.0–44.0, 44.0–88.0] $M_\odot$ we find counterpart-generation probabilities of 4%, 24%, 71%, 100% for the SG model and of 43%, 15%, 31%, 0% for the TQM model.

These results broadly suggest that EM follow-ups should be prioritized for events with larger chirp mass, $\mathcal{M} \geqslant 22\,M_\odot$, which have the highest likelihoods of producing an observable EM counterpart, on timescales of weeks to months after the GW event. For well-localized GW detections that are likely AGN candidates, monitoring of the AGNs within the GW error volume should be conducted at a cadence of ∼every few days to weekly to establish a robust variability baseline. Our results indicate that associated emission may emerge as late as ∼4 months post-GW. This can provide an additional upper limit to counterpart emission timescales.

If we assume that all AGN disks are well represented by the default SG model, then ∼15%–44% of all AGN-channel mergers could have observable EM counterparts, with specific fractions depending on the various model parameters, as detailed in Table 2. Generally, cases where hierarchical mergers are encouraged result in a greater number of observable EM counterparts.

On the other hand, if we assume that all AGN disks are well represented by the TQM model, then ∼0%–40% of all AGN-

**Table 2**
Rate and Percentage of Mergers with Observable EM Flares

| Parameter | Value | SG | | TQM | |
|---|---|---|---|---|---|
| | | $\mathcal{R}_{\rm GW}$ | $f_{\rm EM}$ | $\mathcal{R}_{\rm GW}$ | $f_{\rm EM}$ |
| Default | ⋯ | 11 | 33% | 1 | 20% |
| $\tau_{\rm AGN}$ | 0.25 Myr | 1 | 32% | 0 | 0% |
| | 1 Myr | 8 | 26% | 4 | 22% |
| $R_{\rm outer}$ | $3 \times 10^4 R_g$ | 11 | 15% | ⋯ | ⋯ |
| | $7 \times 10^4 R_g$ | 8 | 33% | ⋯ | ⋯ |
| $\sigma_\mu$ | 0.01 | 8 | 19% | 1 | 28% |
| | 0.1 | 9 | 15% | 1 | 31% |
| | 0.2 | 8 | 19% | 1 | 26% |
| $\gamma$ | $\propto M^{-1}$, [10, 40] $M_\odot$ | 21 | 33% | 2 | 32% |
| | $\propto M^{+2.5}$, [10, 40] $M_\odot$ | 51 | 34% | 8 | 41% |
| | $34 \pm 4\, M_\odot$ | 54 | 32% | 9 | 37% |
| $\sigma_a$ | 0.02 | 11 | 44% | 1 | 35% |
| | 0.20 | 11 | 27% | 1 | 10% |

**Note.** Default model parameter values are $\tau_{\rm AGN} = 0.7$ Myr, $\sigma_\mu = 0$, $\gamma \propto M^{-2}$, [10, 40] $M_\odot$, and $\sigma_a = 0.01$. For each parameter variation we report the BBH merger rate $\mathcal{R}_{\rm GW}$ and the fraction of mergers with observable EM flares, $f_{\rm EM} = n_{\rm flares}/n_{\rm mergers}$, for both SG and TQM disk models.

channel mergers could have observable EM counterparts, with specific fractions depending on the various model parameters, as detailed in Table 2. Generally, 1g mergers dominantly contribute to the fraction of EM bright mergers as a consequence of the lower surface density in the TQM disk model and consequentially diminished hierarchical merger efficiency.

While we have outlined areas of binary parameter space that promote hierarchical mergers (i.e., heavy, high-spin BHs), the specific disk model will critically regulate observable EM counterpart production. In the SG disk, there is a migration trap at the highest-density region that also has a small scale height, and combined these three have the following effects: (i) efficient hierarchical mergers that build heavy, high-spin BHs; (ii) a dense region provides ample gas supply onto the remnant for rapid accretion to drive a powerful jet, especially in the case of a hierarchically formed, heavy BH ($\mathcal{M} \gtrsim 40$); and (iii) a small scale height allows a powerful jet to break out relatively quickly. Nondetection of counterparts may suggest that AGN disks are more extended and/or denser than currently modeled, or that jets in AGN disks cannot be powered for a long enough time to break out. While we suggest longer follow-up to EM counterparts based on our results, almost simultaneous EM counterpart candidates (H. Tagawa et al. 2023) could be explained by geometrical configurations not considered here, especially if the remnant BH is kicked toward the disk photosphere and thus any jet can more promptly emerge.

Because of the dependencies on the disk model, the detection or nondetection of EM counterparts in the upcoming eras of LISA, Cosmic Explorer, and Einstein Telescope will illuminate currently unconstrained elements of the AGN disk environment. LISA will be able to detect heavy EMRIs and light intermediate mass ratio inspirals (IMRIs). Top-heavy BH IMFs explored here efficiently build IMBHs, which are practically guaranteed to produce EM counterparts in our model (see Figure 12). EM counterparts to these mergers are highly promising multimessenger sources. The Cosmic Explorer and Einstein Telescopes will reveal BBH mergers to high redshifts. With these, we expect to be able to robustly estimate the relative contribution to the BBH merger rate from the AGN channel.

From a population perspective, our results identify the conditions that most strongly favor bright jets and potentially observable EM flares. A top-heavy IMF, a moderately long-lived disk, and a substantial local gas reservoir all enhance the production of high-mass, high-spin remnants capable of powering luminous jets. Although dense disks enhance merger rates through gas torques, their optically thick regions may muffle jet signatures: if a jet cannot clear a cavity and promptly penetrate the disk, it may fail to break out. Thus, jet formation and successful escape are tightly regulated by the local disk environment.

In the shock-only scenario, our simulations demonstrated that bolometric shock luminosities are never luminous enough to outshine the AGNs in either disk model. $L_{\rm shock}$ can only outshine a fraction of low-luminosity AGNs and some Seyfert AGNs, especially for large $v_{\rm kick} \sim 10^3$ km s$^{-1}$.

Our estimates for EM counterparts also do not take into account that EM signals could be emitted closer to the photosphere of one side of the disk opposite to the observer and that another fraction of these systems will happen in Type II/edge-on AGNs. More generally, if the EM emission from the jet is highly anisotropic, then the angle between the jet axis and the observer line of sight will further play an important role toward detectability of the emission. Additionally, some binaries in the disk could have an interaction with a third body that perturbs the binary orbit around the SMBH and causes the orbit to become inclined with respect to the disk (and no longer embedded). This population (with a significant nonzero $\chi_p$) could diminish the amount of observable EM counterparts if they push binaries in the migration swamp out into these disk-crossing orbits, and such binaries then merge while orbiting outside of the disk, though this subset of mergers could result in a potentially observable signal via the "delayed jet" mode proposed in J. C. Rodríguez-Ramírez et al. (2025) or the "splash-down" predicted in M. J. Graham et al. (2020). These alternate scenarios are outside of the scope of this paper, but predictive modeling for these events warrants future study.

Even without a detailed EM emission model, our framework yields population-level constraints that identify the most optimistic conditions for observability. Our ongoing work is aimed at predicting both the thermal and nonthermal emission components from jetted remnants to evaluate detectability with specific telescopes. Importantly, the energy of any observable counterpart is bounded by the jet power prescribed by our input physics, providing robust upper limits on EM emission in the case in which mergers always occur at the midplane.[15] Progenitor spin directions, spins, and remnant spins can be inferred through forward modeling, helping to infer plausible jet properties from candidate EM counterparts.

We have explored key uncertainties in the AGN channel and found that the prospects for observable EM emission are primarily controlled by the IMF, the presence of migration traps (and regions of dense gas), and the AGN disk lifetime, while being relatively insensitive to the initial spin distribution and the outer-disk radius. As our modeling improves, these dependencies will enable tighter constraints on the subset of LVK merger remnants in AGN disks capable of producing bright flares and will inform ongoing surveys for EM follow-up to GW events.

[15] We remind the reader that $T_{\rm jet}$ is a free parameter, and thus we will limit $t_{\rm bo}$ and observability.

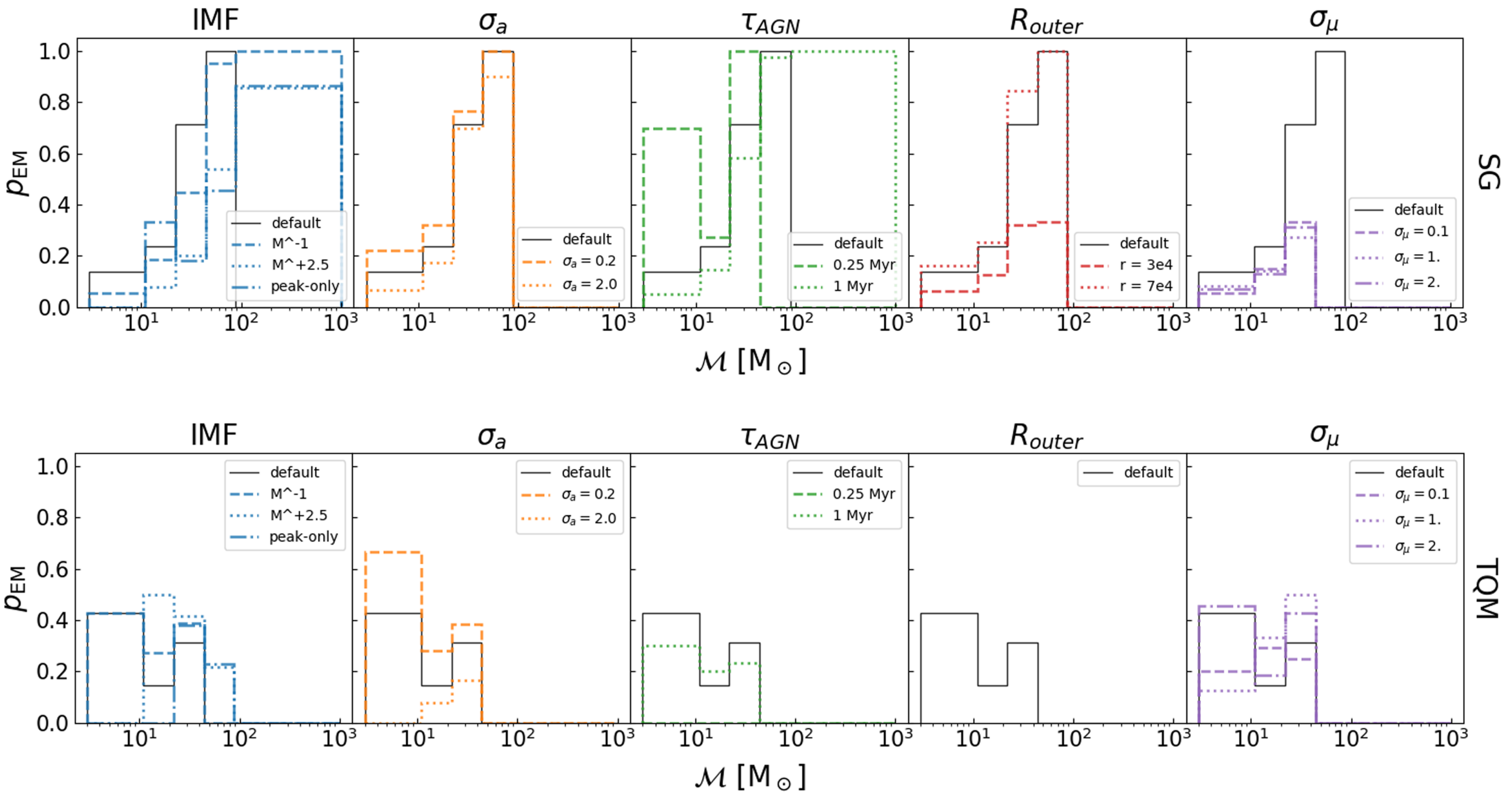

**Figure 12.** Normalized likelihood of producing an observable EM counterpart as a function of chirp mass. For each model parameter explored, we compute the likelihood that a merger produces an observable EM flare within each Grace-DB-like chirp mass bin. Results are shown in separate panels for each varied parameter (from left to right: BH IMF, spin distribution $\sigma_a$, disk lifetime $\tau_{\rm AGN}$, outer-disk radius $R_{\rm outer}$, and turbulence $\sigma_\mu$). Within each panel, different model realizations are color-coded (blue, orange, green, red, purple), while different line styles (see legend) denote individual simulations. All models are compared against the default case, shown as a solid black line. $p_{\rm EM}$ is normalized such that the likelihood is for all observers everywhere in the Universe. A useful rule of thumb for observers is that peak $p_{\rm EM} \sim 0.25$ since only $\mathcal{O}(1/4)$ of EM counterparts should actually be geometrically detectable (assuming that Type I and II AGNs are roughly equal in number—see text). Top row: SG disk model. Bottom row: TQM disk model.

## 6. Conclusions

We have presented the first population synthesis investigation of EM counterparts to BBH mergers in the AGN channel. Using the public, open-source, and reproducible McFACTS code, we have modeled jet and shock luminosities generated by GW-detectable BBH merger remnants embedded in AGN disks. By varying the accretion disk lifetime, turbulence, and radial extent, as well as the initial BH spin distribution and their IMF, we identify regions of parameter space that yield EM luminosities exceeding the host AGN luminosity.

Our simulations demonstrate that accretion-powered jets from remnant BHs can produce significant EM emission, whereas shock-powered emission from kicked remnants remains undetectable in both SG and TQM disk models surrounding a $10^8\ M_\odot$ SMBH.

Our results indicate that denser, SG-like AGN disks preferentially produce high-mass mergers ($\mathcal{M} \gtrsim 44\ M_\odot$) with a very high likelihood of generating an EM counterpart ($p_{\rm EM} \sim 100\%$ for $\mathcal{M} \geqslant 44\ M_\odot$; Figures 10 and 11). SG-like AGN disks are therefore expected to dominate the overall rate of $\geqslant$2g mergers (typically high-mass, high-spin systems) and consequently favor bright EM counterpart production in hierarchically assembled mergers. By contrast, low chirp mass BBH mergers dominate in less dense, TQM-like disks.

Based on these results, we recommend prioritizing EM follow-up of GW events with chirp masses $\gtrsim 22\ M_\odot$ over timescales of up to ~100 days post-merger. Our findings provide quantitative constraints on counterpart rates and can be coupled with detailed emission models to guide future surveys like the Vera C. Rubin Observatory's LSST (Ž. Ivezić et al. 2019). Systematic follow-up, including well-characterized nondetections, will further help constrain AGN disk and NSC properties (K. E. S. Ford & B. McKernan 2026). Upcoming GW catalogs from LVK O4, O5, and other future observing runs will substantially expand the sample available for EM searches, enabling increasingly stringent tests of the AGN-channel scenario.

In future work, we plan to extend the computation of the jet luminosities beyond their purely bolometric values, specifically modeling the thermal and nonthermal radiation from jets and their associated cocoon emission. We will extend this analysis to a synthetic Universe, as in Paper III, in order to explore the impact of varying SMBH mass, NSC mass, and accretion disk model. We are currently applying these methods to generate predictions for LSST and the planned UVEX mission. These efforts will enable increasingly refined forecasts for EM follow-up of GW events in upcoming LVK observing runs, future ground-based GW observatories, and LISA and other space-based GW observatories.

## Data Availability

McFACTS is available at https://github.com/mcfacts, and this work was completed with the version at git hash

`ba3643f`. The data used in this paper is available at Zenodo, DOI:10.5281/zenodo.21705433 (E. McPike et al 2026) .

## Acknowledgments

E.M. would like to thank the CUNY GC MS Astrophysics program and the CCA GW group for support and useful comments. E.M. thanks the CCA at the Flatiron Institute for hospitality while a portion of this research was carried out. The Flatiron Institute is a division of the Simons Foundation. E.M. would like to thank Phil Armitage, Tomás Cabrera, Will Farr, Carrie Filion, Keefe Mittman, and Amy Secunda for helpful conversations and feedback on this work. R.P. acknowledges support by NASA award 80NSSC25K7554. B.M., K.E.S.F., and H.E.C. are supported by NSF AST-2206096. B.M., K.E.S. F., and V.P. are supported by NSF AST-1831415 and Simons Foundation grant 533845. V.D. acknowledges support from the Natural Sciences and Engineering Research Council of Canada (NSERC) under grant RGPIN-2023-05511 and the Research Corporation for Science Advancement's Scialog Initiative under grant No. SALSST-2024-089b with support from the Brinson Foundation. R.O.S. acknowledges support from NSF PHY-2309172 and AST-2206321. J.P. acknowledges support provided by Schmidt Sciences, LLC. V.P. acknowledges support from NASA 24-MSF24-0012. This material is based on work supported by NSF's LIGO Laboratory, which is a major facility fully funded by the National Science Foundation.

*Facilities:* LIGO, PO:1.2m.

*Software:* `Astropy` (Astropy Collaboration et al. 2013, 2018, 2022), matplotlib (J. D. Hunter 2007), `McFACTS` (B. McKernan et al. 2025), numpy (C. R. Harris et al. 2020), `pAGN` (D. Gangardt et al. 2024), pandas (W. McKinney 2010), `precession` (D. Gerosa & M. Kesden 2016).

## Appendix
## Reproducibility

Table 3 contains the parameters we varied in the `McFACTS` code to produce the results shared in this appendix.

**Table 3**
Parameter Choices for Each Run Described Herein

| Run | Parameter | Value |
|---|---|---|
| `sg_tagn_025Myr` | `timestep_num` | 25 |
| `sg_tagn_1Myr` | `timestep_num` | 100 |
| `sg_turb01` | `phenom_turb_std_dev` | 0.01 |
| | `flag_phenom_turb` | 1 |
| `sg_turb1` | `phenom_turb_std_dev` | 0.1 |
| | `flag_phenom_turb` | 1 |
| `sg_turb2` | `phenom_turb_std_dev` | 0.2 |
| | `flag_phenom_turb` | 1 |
| `sg_g1` | `nsc_imf_powerlaw_index` | 1 |
| `sg_g25` | `nsc_imf_powerlaw_index` | 2.5 |
| | `nsc_imf_bh_method` | `power` |
| `sg_gpeak` | `nsc_imf_bh_method` | `peak` |
| `sg_spin_dist_sigma_02` | `nsc_bh_spin_dist_sigma` | 0.02 |
| `sg_spin_dist_sigma_2` | `nsc_bh_spin_dist_sigma` | 0.20 |

**Note.** For a default run, all parameters are listed in Table 1. We list default values under `sg_default` and denote changed values for subsequent runs. To reproduce a specific run, locate the parameter in `model_choice.ini` and alter it to the listed value. For all TQM disk model runs, relevant runs are labeled "`tqm`" instead of "`sg`" and the parameter `disk_model_name` is set to `thompson`.
`disk_model_name` determines the disk model. The default disk model is set to `sirko_goodman`. `nsc_imf_powerlaw_index` determines the slope of the probability distribution in the Pareto distribution function we draw initial masses from (see Paper I). Default IMF slope is set to 2. `nsc_imf_bh_method` changes the method used to populate the initial masses. `flag_use_surrogate` is a choice of –1 and 0, which, respectively, turn on and off precession (V. Delfavero et al. 2025b). Our default model sets this parameter to –1, but the `McFACTS` code has the parameter set to 0. `disk_radius_outer` determines the outermost radius of the accretion disk. The default value is $5 \times 10^4\,R_g$. `flag_phenom_turb` determines whether the disk is considered turbulent. Default is set to 0. With `flag_phenom_turb` set to 1, we add a term to consider stochastic migration due to turbulence, `phenom_turb_std_dev`. `phenom_turb_std_dev` determines how turbulent the disk is; the default value is 0.1. `nsc_bh_spin_dist_sigma` determines the spread of the Gaussian used to draw initial BH spins from. The default value is set to 0.1. The default time step is $10^4$ (see Paper I). `timestep_num` determines the number of time steps in a run, i.e., the lifetime of the accretion disk. The default value is 70 for an accretion disk lifetime of 0.7 Myr.

## ORCID iDs

Emily McPike https://orcid.org/0009-0008-5622-6857
Rosalba Perna https://orcid.org/0000-0002-3635-5677
K.E. Saavik Ford https://orcid.org/0000-0002-5956-851X
Barry McKernan https://orcid.org/0000-0002-9726-0508
Harrison E. Cook https://orcid.org/0000-0001-7163-8712
Vera Delfavero https://orcid.org/0000-0001-7099-765X
Miranda McCarthy https://orcid.org/0009-0005-9964-4790
Kaila Nathaniel https://orcid.org/0000-0003-2430-9515
Jake Postiglione https://orcid.org/0000-0003-0738-8186
Nicolas Posner https://orcid.org/0009-0004-4600-5074
Varun Pritmani https://orcid.org/0009-0000-3666-0586
Shawn Ray https://orcid.org/0009-0005-5038-3171
Richard O'Shaughnessy https://orcid.org/0000-0001-5832-8517